\documentclass[onecolumn,authoryear]{els-mrw} 

\usepackage{amsmath,amssymb,amsfonts,amsthm,makeidx,graphicx}
\usepackage{txfonts}
\usepackage{helvet}

\begin{document}

\chapter{Quasi-real photons as a probe of exotic nuclei}\label{chap1}

\author[1]{Takashi Nakamura}%
\author[2,3]{Carlos A. Bertulani}%

\address[1]{\orgname{Institute of Science Tokyo}, \orgdiv{Department of Physics}, \orgaddress{2-12-1 O-Okayama, Meguro, Tokyo, 152-8550, Japan}}
\address[2]{\orgname{East Texas A\&M University}, \orgdiv{Department of Physics and Astronomy}, \orgaddress{Commerce, Texas 75429, USA}}
\address[3]{\orgname{Technische Universit\"{a}t Darmstadt}, \orgdiv{ Institut f\"{u}r Kernphysik}, \orgaddress{64289, Darmstadt, Germany}}

\articletag{Chapter Article tagline: Reprint.}

\maketitle

%\begin{glossary}[Glossary]
%\term{Europe} the model is a coherent view of capital markets data that allows users to interact with the content in a consistent manner.

%\term{Primates} regardless of the source. Essentially, of sources. Properly deployed.

%\end{glossary}

%\begin{glossary}[Nomenclature]
%\begin{tabular}{@{}lp{34pc}@{}}
%AF &Assessment Factor\\
%ECHA &European Chemical Agency\\
%EPM &Equilibrium Partitioning Method Equilibrium Partitioning Method Equilibrium Partitioning Method Equilibrium\hfill\break Partitioning Method\\
%ERA &Ecological Risk Assessment\\
%HC &Hazardous Concentration\\
%\end{tabular}
%\end{glossary}

%\documentclass[english,aps,prc,0superscriptaddress,longbibliography,notitlepage]{revtex4-1}
%%\documentclass[prl,aps,twocolumn,groupedaddress,showpacs,superscriptaddress,longbibliography]{revtex4-1}
%\usepackage{amsmath,amssymb,bm}
%\usepackage{soul}
%\usepackage{epsfig}
%\usepackage{graphicx}
%\usepackage{amsmath}
%\usepackage{lipsum}
%\usepackage{color}
\makeatother

\newcommand{\red}[1]{\textcolor{red}{#1}}

% Optional: configure link appearance (removes colored boxes around links)
%\hypersetup{
%    colorlinks=true,
%    linkcolor=blue,
%    filecolor=magenta,      
%    urlcolor=cyan,
%    citecolor=blue,
%    pdftitle={Your Title},
%    pdfpagemode=FullScreen,
%    }
    
\newcommand{\etal}{{\it et al.,\;}}
\newcommand{\avg}[1]{\langle #1 \rangle}
\newcommand{\eqn}[1]{\rm{Eq.}~(\ref{#1})}
\newcommand{\fig}[1]{\rm{Fig.}~\ref{#1}}
\newcommand{\apriori}{\textit{a~priori} }
\newcommand{\pictsize}{0.98}
\newcommand{\beq}{\begin{equation}}
\newcommand{\eeq}{\end{equation}}

\newcommand{\be}{\begin{equation}}
\newcommand{\ee}{\end{equation}}
\newcommand{\ba}{\begin{eqnarray}}
\newcommand{\ea}{\end{eqnarray}}
\newcommand{\bea}{\begin{eqnarray}}
\newcommand{\eea}{\end{eqnarray}}

%---nakamura
\newcommand{\erel}{E_{\rm rel}}
\newcommand{\ex}{E_{\rm x}}
\newcommand{\eone}{E\mathrm{1}}
\newcommand{\mone}{M\mathrm{1}}

\newcommand{\beone}{B(\eone)}
\newcommand{\dsderel}{\frac{d\sigma}{dE_{\rm rel}}}
\newcommand{\neone}{n_{\eone}(E_{\rm x})}
\newcommand{\sn}{S_{n}}
\newcommand{\snn}{S_{2n}}
\newcommand{\dbderel}{\frac{dB(\eone)}{dE_{\rm rel}}}
\newcommand{\dbde}{dB(\eone)/dE_{\rm rel}}
\newcommand{\dbdest}{dB(\eone)/dE^*}
\newcommand{\dsderelone}{d\sigma_{\rm CB}/dE_{\rm rel}}
\newcommand{\vecrone}{\mathbf{r_1}}
\newcommand{\vecrtwo}{\mathbf{r_2}}
\newcommand{\rcn}{\mathbf{r}_{cn}}
\newcommand{\rcnn}{\mathbf{r}_{c-nn}}
\newcommand{\rnn}{\mathbf{r}_{nn}}
\newcommand{\vecbeam}{\mbox{\boldmath$P$}(^{11}{\rm Be})}
\newcommand{\vecfrag}{\mbox{\boldmath$P$}(^{10}{\rm Be})}
\newcommand{\vecneut}{\mbox{\boldmath$P$}(n)}
\newcommand{\dsdomegade}{\frac{d\sigma^2}{d\Omega d\erel}}
\newcommand{\dsdomega}{\frac{d\sigma(\theta,\ex)}{d\Omega}}
\newcommand{\dndomega}{\frac{dn_{\rm E1}(\theta,\ex)}{d\Omega}}
\newcommand{\egam}{E_{\gamma}}
\newcommand{\shalfzero}{{}^{10}\mathrm{Be}(0_1^+)\otimes 1s_{1/2}}
\newcommand{\dhalftwo} {{}^{10}\mathrm{Be}(2_1^+)\otimes 0d_{5/2}}
\newcommand{\sigmac}{\sigma(\eone)}
\newcommand{\dbdex}{\frac{dB(\eone)}{d\ex}}
\newcommand{\sigmaonenpb}{\sigma_{-1n}({\rm Pb})}
\newcommand{\sigmaonencb}{\sigma_{-1n}({\rm C})}
\newcommand{\sigmaonenc}{\sigma_{-1n}(\eone)}
\newcommand{\sigmaeonegs}{\sigma(\eone; 0^+_{\rm gs})}
\newcommand{\sigmacgs}{\sigma(C; 0^+_{\rm gs})}
\newcommand{\pconf}{{}^{30}\mathrm{Ne}(0_1^+)\otimes 1p_{3/2}}
\newcommand{\fconf}{{}^{30}{\mathrm{Ne}(0_1^+)}\otimes 0f_{7/2}}
\newcommand{\cs}{C^2S}
\newcommand{\jpi}{J^\pi}

%\bibliographystyle{apsrev4-1}

%\begin{document}

%\title{Quasi-real photons as a probe of exotic nuclei}
%\author{Takashi Nakamura}\email{nakamura@phys.sci.isct.ac.jp}
%\affiliation{Department of Physics, Institute of Science Tokyo, 
%2-12-1 O-Okayama, Meguro, Tokyo, 152-8550, Japan}

%\author{Carlos A. Bertulani}\email{Carlos.Bertulani@etamu.edu}
%\affiliation{Department of Physics and Astronomy, 
%East Texas A\&M University, Commerce, Texas 75429, USA}
%\affiliation{
%Technische Universit\"{a}t Darmstadt, Institut f\"{u}r 
%Kernphysik, 64289, Darmstadt, Germany}


%\date{\today}

%\maketitle

\section{How can we probe a nucleus only $\sim 10^{-14}$ m in size}
As early as 1904, Nagaoka proposed what became known as the Saturnian model of the atom [\cite{Nag04}]. Inspired by the structure of Saturn and its rings, he suggested that the atom consisted of a massive, positively charged central core, with electrons orbiting it in a ring-like configuration, and that the central core contained 
most of the atomic mass.
This represented a radical departure from J. J. Thomson’s ``plum pudding" model, in which positive charge was thought to be distributed uniformly throughout the atom [\cite{Thom04}]. Nagaoka instead assumed a highly concentrated positive charge, conceptually similar to what would later be called the nucleus.
However, his model faced stability problems under classical electrodynamics, and at the time there was no experimental evidence supporting such a concentrated charge distribution.

In 1909, Hans Geiger and Ernest Marsden, working under the direction of Ernest Rutherford at the University of Manchester, performed one of the most important experiments in physics: the scattering of alpha particles from a gold foil. Their results overturned the prevailing ``plum pudding'' model of the atom and led to the discovery of the atomic nucleus
[\cite{Rut11,GM09,GM13}].
At the time, J.~J.~Thomson's model described the atom as a diffuse sphere of positive charge with electrons embedded within it. If this model were correct, positively charged alpha particles---helium nuclei---fired at a thin foil would experience only small deflections as they passed through its weak, distributed electric field. 

In the experiment, alpha particles from a radioactive source were directed at an extremely thin gold foil. A movable zinc sulfide screen detected the scattered particles through tiny flashes of light, or scintillations. Most of the alpha particles passed through the foil with little or no deflection, indicating that atoms are largely empty space. However, a small fraction were deflected through large angles, and a very small number were scattered almost directly backward.
Such large-angle deflections were incompatible with a diffuse charge distribution. 

Rutherford interpreted the results using Coulomb's law, proposing that nearly all the atom's positive charge and mass are concentrated in a tiny central region: the nucleus. The strong electric repulsion between the positively charged alpha particle and this compact nucleus could explain the observed scattering [\cite{Rut11,GM09,GM13}].
He derived the classical scattering formula
\begin{equation}
\frac{d\sigma}{d\Omega}
=
\left(
\frac{Z z e^2}{16\pi \varepsilon_0 E}
\right)^2
\frac{1}{\sin^4(\theta/2)},
\end{equation}
where $ze$ and $Ze$ are the electric charges of the projectile particle (for $\alpha$, $z=2$) and the target (for Au, $Z=79$), respectively, and $E$ is the energy of the projectile. This formula predicts a characteristic $\sin^{-4}(\theta/2)$ dependence on the scattering angle $\theta$. Experimental measurements confirmed this angular distribution, providing firm evidence for the nuclear model.

From these results, Rutherford concluded that the atom consists of a tiny, massive nucleus, about $10^{-14}\,\mathrm{m}$ in size, surrounded by electrons. Unlike Nagaoka’s theoretical proposal, Rutherford’s model was based on quantitative scattering data and was supported by a predictive formula for Coulomb scattering. This model replaced Thomson's picture and laid the foundation for Niels Bohr's quantum theory of the atom~[\cite{Bohr1913}], the later quantum-mechanical description of atomic structure developed by Schr\"odinger~[\cite{Schrodinger1926}],  and the development of nuclear physics, including James Chadwick's discovery of the neutron in 1932~[\cite{Chad1932,Chad1932Existence}].

Beyond revealing the nuclear structure of matter, the experiment established scattering as a powerful method for probing microscopic systems. From electron scattering to modern high-energy collisions, the same basic principle---inferring internal structure from angular distributions---remains central to physics. The Rutherford experiment thus marks the birth of the nuclear atom and one of the great turning points in science.

\begin{figure}[tb]
\begin{center}
{\includegraphics[width=8cm]{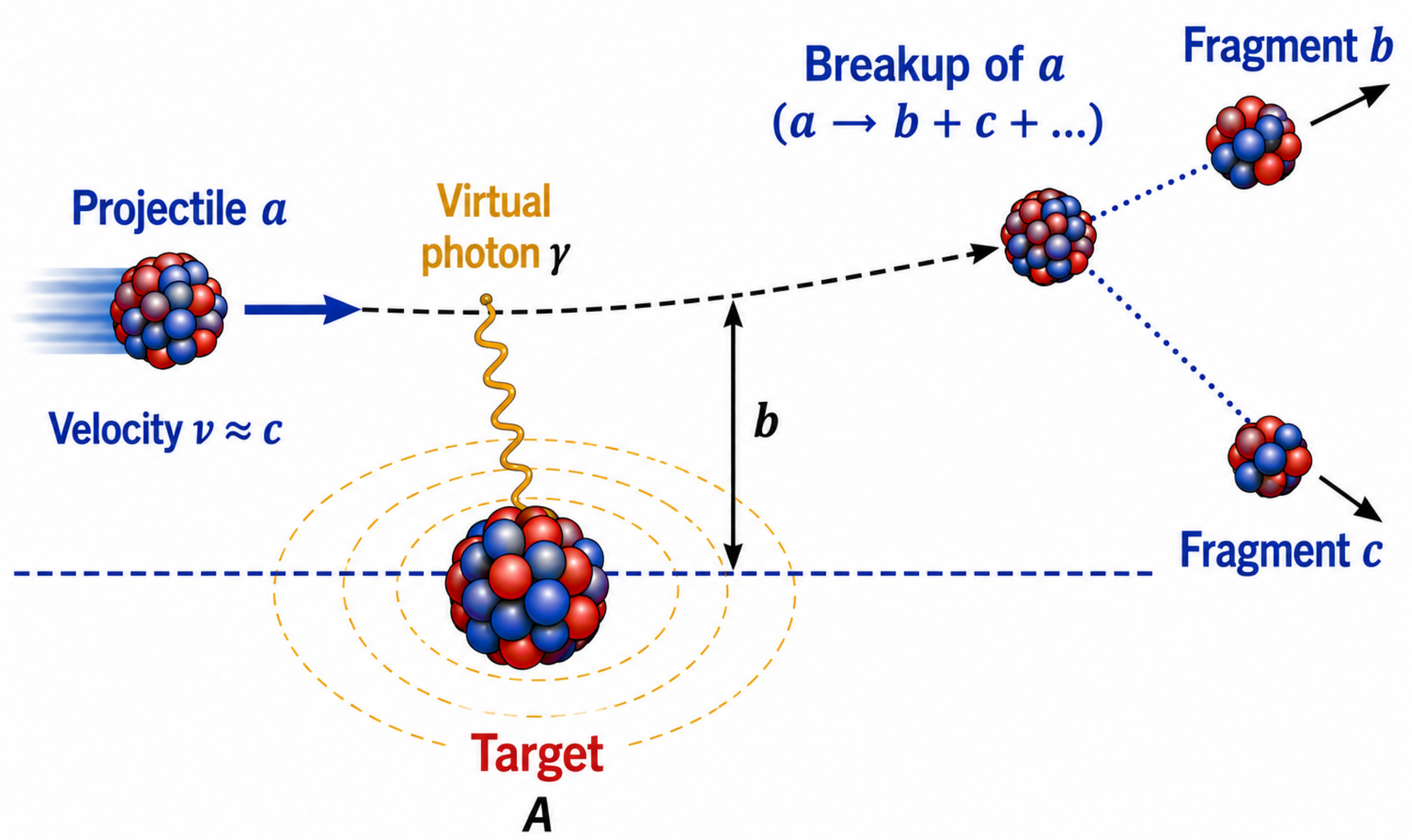}}
\end{center}
\vspace{-0.5cm}
\caption{\label{cdfig}
Schematic view of Coulomb excitation followed by the dissociation of a fast nuclear projectile incident on a large-Z nuclear target.}
\end{figure}

In this article, we discuss a modern form of Rutherford scattering: Coulomb excitation at relativistic energies, with particular emphasis on Coulomb dissociation, in which an excited projectile subsequently breaks up, 
as schematically illustrated in Fig.~\ref{cdfig}.
Rutherford scattering usually refers to elastic scattering induced by the Coulomb interaction between the projectile and the target, in which both particles remain in their ground states throughout the process. In contrast, Coulomb excitation is an inelastic process in which either the target or the projectile is excited by the Coulomb interaction, which can equivalently be described as the absorption of a virtual photon.

The article is organized as follows. Section~\ref{sec:virtualphoton} discusses 
Coulomb excitation at relativistic energies. 
Section~\ref{sec:RIbeam} provides an overview of 
Coulomb excitation and Coulomb dissociation using rare-isotope beams.
The subsequent sections are devoted to applications of 
Coulomb dissociation using rare-isotope beams. 
Section~\ref{sec:halo} discusses the Coulomb dissociation 
of neutron-halo nuclei, and Section~\ref{sec:astro} reviews the application of Coulomb dissociation to the determination of radiative-capture reaction rates 
relevant to stellar reactions.
Finally, Section~\ref{sec:summary} summarizes the article and presents 
perspectives for Coulomb dissociation studies in the near future.

%\section{How to excite and vibrate a nucleus \label{sec:virtualphoton}}
\section{How can we excite and vibrate a nucleus \label{sec:virtualphoton}}

Since the 1950s, heavy-ion projectiles have been used as sources of time-dependent electromagnetic fields that act as effective quasi-real photon probes. It has been shown that Coulomb excitation can serve as an alternative to ``real" photon absorption. In modern rare-isotope beam experiments, heavy, high-$Z$ targets are used to excite and dissociate nuclear projectiles in flight, 
as shown in Fig.\ref{cdfig}. This approach has opened a new era in nuclear physics because most exotic and neutron-rich nuclei can be studied only as secondary rare-isotope beams incident on fixed targets.

In the decades following the Rutherford experiment, the inelastic scattering of heavy ions became an important tool for investigating the structure of
both stable and unstable nuclei. The angular distributions of inelastically scattered fragments are particularly useful for
unambiguously identifying the multipolarity of the interaction and
consequently the spins and parities of the excited states. To correctly describe these angular distributions, absorption and diffraction effects must be
properly taken into account~[\cite{BB88_8,BN93-6}].

The double-differential cross section for Coulomb excitation 
in high-energy nucleus-nucleus collisions is given by
\begin{equation}
{\frac{d^{2}\sigma_{C}}{d\Omega\,dE_{\gamma}}} \left(  E_{\gamma}\right)
={\frac{1}{E_{\gamma}}} \sum_{\pi\lambda}{\frac{dn_{\pi\lambda}}{d\Omega}%
} \sigma_{\gamma}^{\pi\lambda} \left(  E_{\gamma}\right) \label{dsig},
\end{equation}
where $E_\gamma = \hbar \omega$ is the excitation energy, $\sigma_{\gamma}^{\pi\lambda}\left(  E_{\gamma}\right)  $ is the
photonuclear cross section for the absorption of a real photon with energy
$E_{\gamma}$, and $dn_{\pi\lambda}/d\Omega$ is the differential 
\textit{virtual photon number}, given by [\cite{BN93-6}]
\begin{equation}
{\frac{dn_{\pi\lambda}}{d\Omega}}=Z_{1}^{2}\alpha \left(  {\frac{\omega
k}{\gamma v}}\right)  ^{2}{\frac{\lambda\left[  (2\lambda+1)!!\right]  ^{2}%
}{(2\pi)^{3} (\lambda+1)}} \sum_{m} |G_{\pi\lambda m}|^{2}(c/v) |\Omega
_{m}(q)|^{2}\;.\label{dn}%
\end{equation}
Here, $|G_{\pi\lambda m}|^{2}(c/v)$ denotes the Winther-Alder function for an ion moving with velocity $v$~[\cite{WA79}], $\alpha=e^{2}/\hbar c$, and [\cite{BN93-6}]
\begin{equation}
\Omega_{m}(q)=\int_{0}^{\infty}db\ b\ J_{m}(qb)\ K_{m}\left(  {\frac{\omega
b}{\gamma v}}\right)  \ \exp\left\{  i\chi(b)\right\}  \;,\label{Omeg}%
\end{equation}
where $q=2k\sin(\theta/2)$ is the momentum transfer, with $\theta$ being 
%and $\phi$ are
the
%and azimuthal 
scattering angle. The eikonal phase shift $\chi(b)$
accounts for the effects of quantum scattering and absorption associated with 
the optical potential. 

The photoabsorption cross section entering Eq.~(\ref{dsig}) can be expressed in
terms of the reduced electromagnetic transition probability as
\begin{equation}
\sigma_{\gamma}^{\pi\lambda}(E_\gamma)
=
\frac{(2\pi)^3(\lambda+1)}
{\lambda\left[(2\lambda+1)!!\right]^2}
\left(\frac{E_\gamma}{\hbar c}\right)^{2\lambda-1}
\frac{dB(\pi\lambda;E_\gamma)}{dE_\gamma},
\label{sigmagamma}
\end{equation}
where $dB(\pi\lambda;E_\gamma)/dE_\gamma$ is the reduced transition strength
distribution. Equations~(\ref{dsig})
and (\ref{sigmagamma}) relate the measured Coulomb
dissociation cross section to the electromagnetic response of the nucleus and
are valid for both discrete and continuum excitations. For a discrete state of
excitation energy $E_x$, one replaces
$dB(\pi\lambda;E_\gamma)/dE_\gamma
\rightarrow
B(\pi\lambda)\delta(E_\gamma-E_x)$.

The angle-integrated cross section for Coulomb excitation can be obtained
from Eqs.~(\ref{dsig}) and (\ref{dn}), using the approximation
$d\Omega\simeq2\pi qdq/k^{2}$, which is valid at small scattering angles
and for small energy losses. Using the closure relation for the Bessel
functions, one obtains [\cite{BB88_8}]
\begin{equation}
{\frac{d\sigma_{C}}{dE_{\gamma}}} \left(  E_{\gamma}\right)  ={\frac
{1}{E_{\gamma}}} \sum_{\pi\lambda}n_{\pi\lambda}\left(  E_{\gamma}\right)
\ \sigma_{\gamma}^{\pi\lambda} \left(  E_{\gamma}\right)  \;,\label{sig}%
\end{equation}
where the total number of virtual photons with energy $\hbar\omega$ is given
by
\begin{equation}
n_{\pi\lambda}(\omega)=Z_{1}^{2}\alpha {\frac{\lambda\left[  (2\lambda
+1)!!\right]  ^{2}}{(2\pi)^{3}\ (\lambda+1)}} \sum_{m}\ |G_{\pi\lambda
m}(c/v)|^{2}\ g_{m}(\omega)\;,\label{n}%
\end{equation}
with
\begin{equation}
g_{m}(\omega)=2\pi \left(  {\frac{\omega}{\gamma v}}\right)  ^{2}\ \int
db\ b K_{m}^{2}\left(  {\frac{\omega b}{\gamma v}}\right)   \exp\left\{
-2 \chi_{I}(b)\right\}  \;,\label{g}%
\end{equation}
where $\chi_{I}(b)$ is the imaginary part of the eikonal phase $\chi(b)$.

It is common to assume a sharp-cutoff for pure relativistic Coulomb excitation so that  $\chi_{I}(b)=\Theta(b-b_{min})$. In this case, using $\beta = {v}/{c},$ and
$\gamma = {1}/{\sqrt{1-\beta^2}},$ and the adiabaticity parameter
$\xi ={\omega b_{\min}}/{\gamma v},$ where $\hbar\omega$ is the photon energy and $b_{\min}$ denotes the minimum
impact parameter retained in the electromagnetic collision. In the Bertulani-Baur convention [\cite{BB88_8}], the equivalent-photon numbers integrated over
$b>b_{\min}$ are written in terms of the modified Bessel functions
$K_\nu(\xi)$.
For electric dipole excitation one obtains
\begin{equation}
n_{E1}(\omega)
=
\frac{2 Z_T^2 \alpha}{\pi\beta^2}
\left[
\xi K_0(\xi)K_1(\xi)
-
\frac{\beta^2\xi^2}{2}
\left(
K_1^2(\xi)-K_0^2(\xi)
\right)
\right],
\label{eq:nE1_BB}
\end{equation}
where $Z_T$ is the charge of the nucleus producing the electromagnetic
field and $\alpha=e^2/(\hbar c)$ is the fine-structure constant.

The corresponding electric-quadrupole photon number is
\begin{align}
n_{E2}(\omega)
=
\frac{2 Z_T^2\alpha}{\pi\beta^4}
\Bigg\{
2(1-\beta^2)K_1^2(\xi)
+
\xi(2-\beta^2)^2 K_0(\xi)K_1(\xi)
-
\frac{\beta^4\xi^2}{2}
\left[
K_1^2(\xi)-K_0^2(\xi)
\right]
\Bigg\}.
\label{eq:nE2_BB}
\end{align}
The additional factor $1/\beta^4$, compared with the dipole case, reflects
the stronger velocity dependence of the electric quadrupole field at
nonrelativistic and intermediate energies.

For magnetic dipole excitation the corresponding expression is
\begin{equation}
n_{\mone}(\omega)
=
\frac{2 Z_T^2\alpha}{\pi}
\left[
\xi K_0(\xi)K_1(\xi)
-
\frac{\xi^2}{2}
\left(
K_1^2(\xi)-K_0^2(\xi)
\right)
\right].
\label{eq:nM1_BB}
\end{equation}
Thus, in the ultrarelativistic limit, $\beta\rightarrow 1$ and
$\gamma\gg 1$, the $\eone$, $E2$, and $\mone$ spectra approach one another,
as expected because the Lorentz-contracted electromagnetic field increasingly
resembles a pulse of transverse radiation.

For very high beam energies, where $\beta\simeq1$, the dominant dependence
of the photon spectrum is governed by
$
\xi \simeq {\omega b_{\min}}/{\gamma c}.
$
The exponential decrease of $K_\nu(\xi)$ for $\xi\gtrsim1$ produces the
well-known adiabatic cutoff,
$
\hbar\omega_{\max}
\sim
{\gamma\hbar v}/{b_{\min}},
$
showing explicitly that increasing the Lorentz factor extends the equivalent
photon spectrum toward higher excitation energies. At sufficiently large
$\gamma$, all electromagnetic multipoles tend toward a common transverse
photon spectrum, whereas at intermediate energies the differences between
$E1$, $E2$, and $M1$ photon numbers can remain substantial.
The sharp-cutoff prescription $b>b_{\min}$ is useful analytically, but in
realistic calculations it may be replaced by a smooth absorption probability
or survival factor,
\begin{equation}
n_{\pi\lambda}(\omega)
=
2\pi
\int_0^\infty b\,db\,
P_{\rm surv}(b)\,
N_{\pi\lambda}(\omega,b),
\label{eq:smooth_equiv_photons}
\end{equation}
where $P_{\rm surv}(b)$ suppresses trajectories for which strong nuclear
absorption occurs. This form is particularly useful when Coulomb dissociation
is combined with Glauber, optical-model, or eikonal descriptions of the
nuclear interaction.
The standard derivation and a detailed discussion of these relativistic
equivalent-photon spectra can be found in Refs.~[\cite{BB88_8,BN93-6}].

\begin{figure}[tb]
\begin{center}
{\includegraphics[width=8cm]{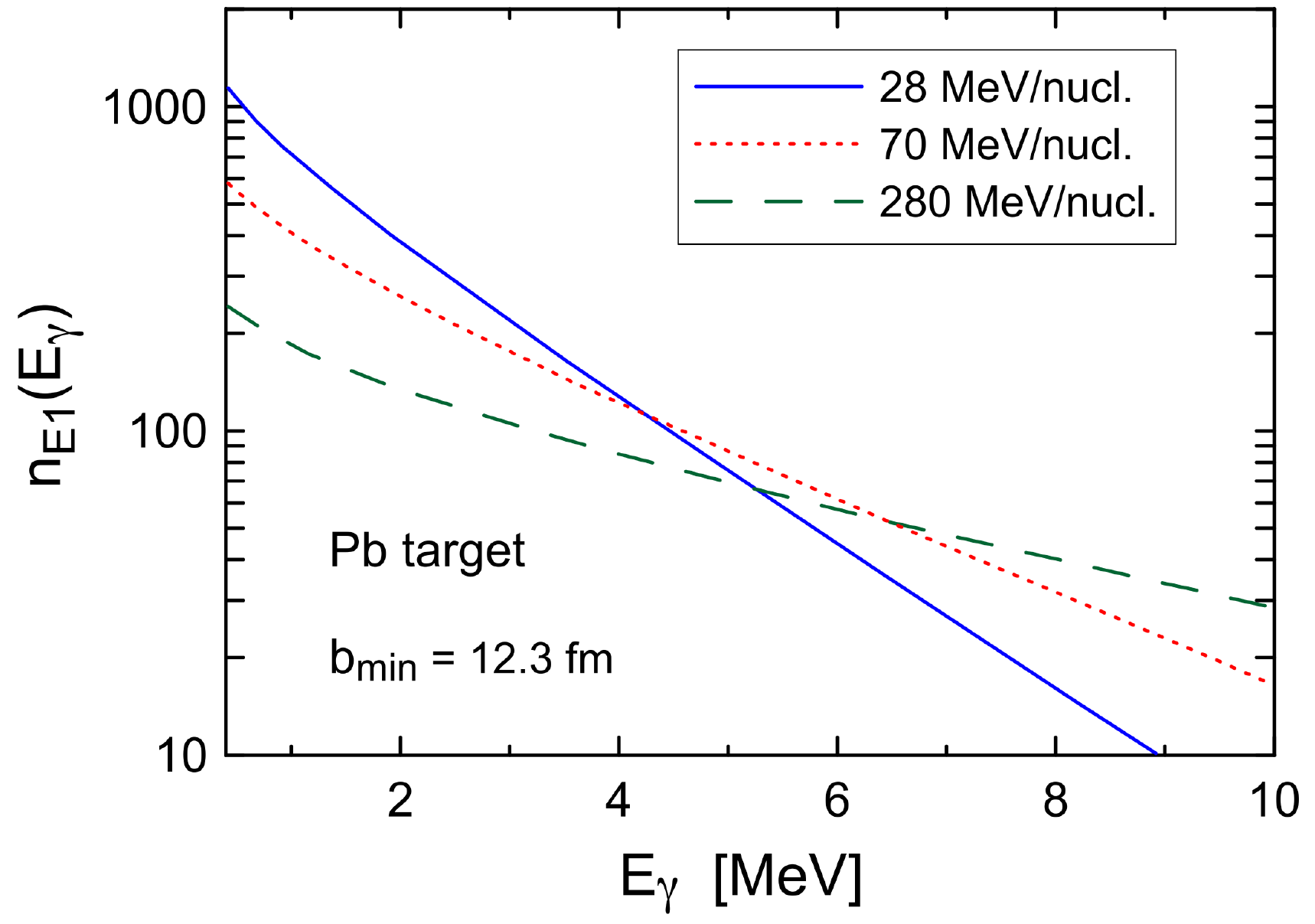}}
\end{center}
\vspace{-0.5cm}
\caption{\label{vpnfig}
Number of virtual photons for the E1 multipolarity, as ``seen'' by a projectile flying by a lead target at impact parameters $b>  12.3$ fm, at three projectile energies.}
\end{figure}

Figure \ref{vpnfig} shows the number of virtual photons for the $\eone$ 
multipolarity, with $E_\gamma =\hbar \omega$, as ``seen'' by a  
projectile passing a lead target at impact parameters  $b>12.3$ fm for three different projectile energies. As the bombarding energy increases, virtual photons with higher energies become available. 
%The number of states accessed in the excitation process is concomitantly increased, which means that high-lying states, such as giant resonances, $E_{GR} \sim 10-20$ MeV, can be excited by the Coulomb field at higher incident energies.
Consequently, a wider range of excited states become accessible, allowing high-lying states such as giant resonances, with $E_{GR}\sim 10–20$~MeV, to be excited by the Coulomb field at higher incident energies.

A straightforward application of probing nuclei with quasi-real photons is the Coulomb excitation of giant resonances.
The corresponding photonuclear cross sections, $\sigma_{\gamma}^{GR}$, can be parametrized by a Lorentzian form,
\begin{equation}
\sigma_{\gamma}^{GR}(E)=\sigma_{0} \frac{E^{2}\Gamma^{2}}{(E^{2}-E_{GR}^{2})^{2}+E^{2}\Gamma^{2}},
\end{equation}
which can be applied to the excitation of giant dipole resonances (GDR), giant quadrupole resonances (GQR) and other giant resonances, with the 
well-known strength $\sigma_0$ and width $\Gamma$ parameters [\cite{AUMANN1996321}]. 

\begin{figure}
\begin{center}
{\includegraphics[width=7cm]{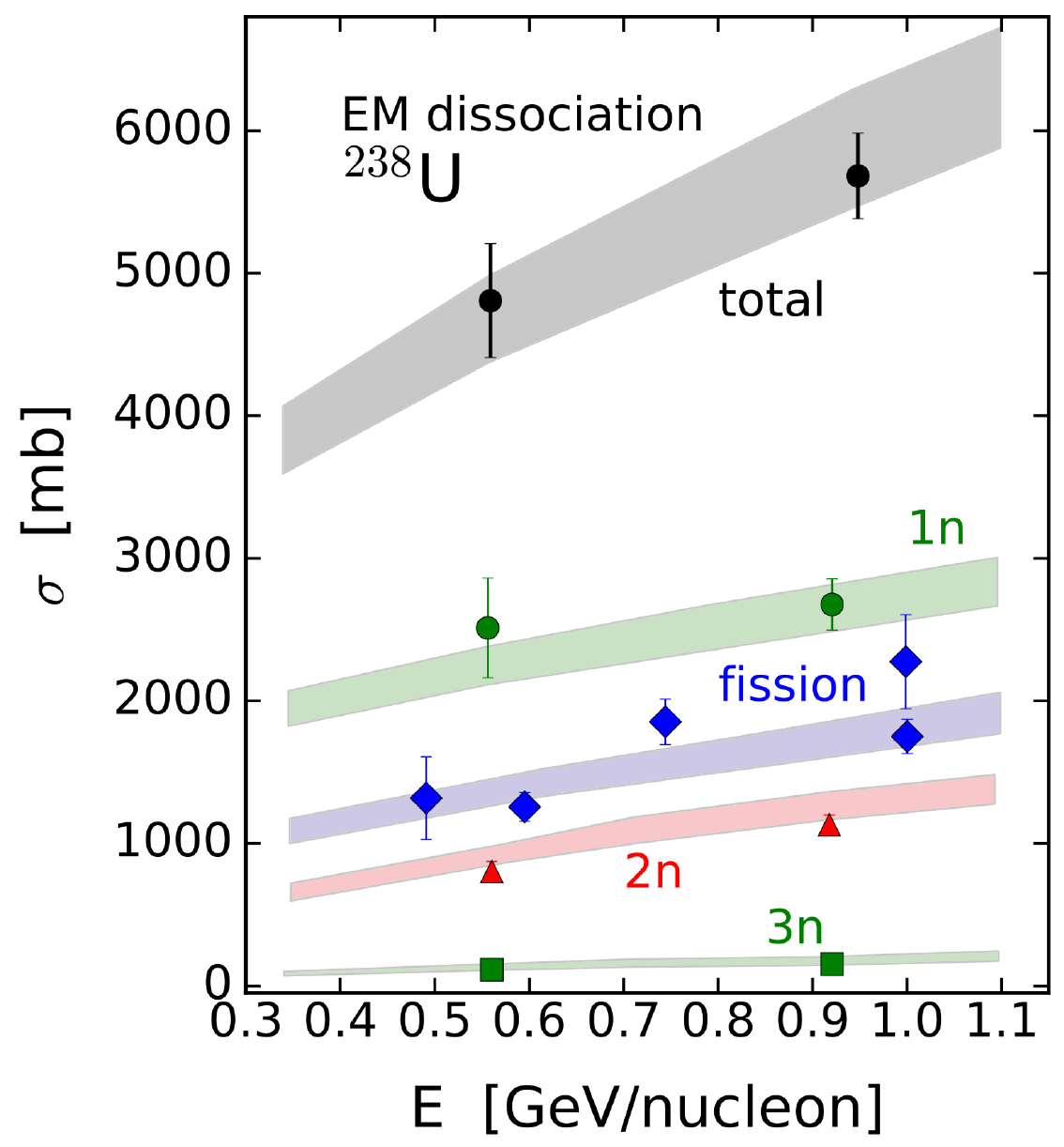}}
\end{center}
\vspace{-0.5cm}
\caption{\label{emdis}
Cross section for electromagnetic dissociation of $^{238}$U. The Coulomb fission cross sections (diamonds), measured with $^{208}$Pb beams on $^{238}$U targets, as well as the multiple neutron decay channels, xn, and fission cross sections obtained in inverse kinematics with $^{238}$U beams. The data obtained with Au targets were scaled to the data obtained with Pb targets. The curves show theoretical calculations using two sets of experimental GDR parameters as an input  [\cite{AUMANN1996321}]. }
\end{figure}

It was earlier recognized [\cite{BB88_8}] that the electromagnetic excitation of nuclei in relativistic heavy-ion collisions leads to large cross sections because of the excitation of giant resonances. The decay of these resonances, including the double giant dipole resonance (DGDR), leads to several decay channels, such as $x$n evaporation and fission. This was demonstrated by comparison of the experiment to the theory in Ref.~[\cite{AUMANN1996321}].  Therefore, EM dissociation leads to large interaction cross sections, which can be easily measured in inverse kinematics. In Figure  \ref{emdis} we show the cross section for electromagnetic dissociation of $^{238}$U [\cite{AUMANN1996321}].  The curves show theoretical calculations using two sets of experimental GDR parameters as input. It is clear that,  except for light targets, one should expect that EM dissociation will always be relevant in any nucleus-nucleus collision at relativistic energies and 
%can either be used as a spectroscopic tool or constitute a background.
might be either used as a spectroscopic tool, or become a background for processes where the effects of the strong interaction are studied. 

The first-order equivalent photon method (EPM) provides an excellent description of Coulomb dissociation for sufficiently peripheral collisions at intermediate and relativistic beam energies. In this approximation, the breakup probability is proportional to the photoabsorption cross section, allowing one to extract radiative capture cross sections through the principle of detailed balance. However, as the electromagnetic interaction strength increases or the projectile binding energy becomes very small, higher-order electromagnetic processes become increasingly important. These effects include multiple virtual-photon exchange, continuum-continuum couplings, post-acceleration of charged fragments, and higher-order excitation amplitudes that cannot be represented by a single-photon transition. Their importance depends on the beam energy, the projectile charge, the target charge, and the impact parameter. While first-order perturbation theory is often adequate for loosely bound neutron-rich nuclei at several hundred MeV per nucleon, corrections of several percent up to tens of percent may arise for heavy targets or for proton-rich systems where the Coulomb interaction is considerably stronger.

A rigorous treatment of higher-order effects is naturally provided by solving the time-dependent Schr\"odinger equation or by employing the continuum-discretized coupled-channels (CDCC) method [\cite{Bertulani2003,Esbensen1996,Typel2001,OGATA06,Moro2006,Nunes1996,Capel2004}].  In these approaches, the projectile wave function evolves in the time-dependent electromagnetic field of the target, allowing repeated transitions between bound and continuum states as well as continuum--continuum couplings. Such calculations have demonstrated that higher-order effects generally reduce the extracted electric dipole strength compared with first-order calculations and can modify both energy spectra and angular distributions of the emitted fragments. The magnitude of these corrections depends sensitively on the adiabaticity parameter,
$\xi={\omega b}/{\gamma v}$. 
When $\xi \ll 1$, the collision is sudden and higher-order couplings become more significant, whereas for $\xi \gg 1$ the interaction becomes increasingly adiabatic and the excitation probability is suppressed.

An equally important ingredient in the theoretical description of Coulomb dissociation is the treatment of final-state interactions (FSI) between the breakup fragments. After dissociation, the fragments continue to interact through both nuclear and Coulomb forces, modifying the continuum wave function relative to a simple plane-wave description. In neutron-halo nuclei such as $^{11}$Be and $^{15}$C, neutron--core interactions generate resonant and non-resonant continuum structures that strongly influence the observed energy distributions. In proton-rich nuclei, the Coulomb repulsion between the emitted proton and the residual core introduces additional distortions that affect the shape and magnitude of the breakup spectrum, particularly near threshold where Coulomb wave functions differ substantially from free-particle solutions. Consequently, realistic continuum wave functions generated from optical-model or microscopic interactions are required for quantitative analyses [\cite{Bertulani2003,Esbensen1996,Typel2001,Yahiro2012,Moro2006,Nunes1996,Capel2004}].
The interplay between higher-order electromagnetic effects and final-state interactions has been extensively investigated for benchmark systems such as $^{8}$B, $^{11}$Be, and $^{15}$C. Comparisons between first-order perturbation theory, dynamical calculations, CDCC analyses, and time-dependent approaches have shown that modern theoretical uncertainties can generally be reduced to a few percent for carefully selected kinematic conditions. 

Although electric dipole ($E1$) transitions dominate the Coulomb dissociation of most weakly bound nuclei, particularly neutron-halo systems, higher electromagnetic multipolarities may contribute significantly under appropriate kinematic conditions. Electric quadrupole ($E2$) excitations become increasingly important for heavier projectiles, proton-rich nuclei, larger excitation energies, or collisions involving smaller impact parameters where the electromagnetic field is stronger. Likewise, magnetic dipole ($M1$) transitions may contribute when low-lying spin-flip states are accessible or when the projectile possesses significant magnetic transition strength. For many astrophysical applications, such as the extraction of the $^{7}$Be$(p,\gamma)^{8}$B capture cross section, accurate determination of the small $E2$ contribution is essential because it may influence the inferred astrophysical $S$ factor at the level of several percent. Modern analyses therefore include realistic calculations of all relevant multipole amplitudes together with their interference.

Another important consideration is the interference between Coulomb and nuclear breakup amplitudes. Although Coulomb dissociation experiments are designed to employ sufficiently large impact parameters so that the interaction is predominantly electromagnetic, the nuclear interaction cannot be completely eliminated experimentally. The measured breakup amplitude may therefore be written schematically as $
\mathcal{M}
=
\mathcal{M}_{\rm C}
+
\mathcal{M}_{\rm N},
$
leading to the differential cross section
$
|\mathcal{M}|^2
=
|\mathcal{M}_{\rm C}|^2
+
|\mathcal{M}_{\rm N}|^2
+
2\,{\rm Re}
\left(
\mathcal{M}_{\rm C}
\mathcal{M}_{\rm N}^{*}
\right).
$
The interference term may either increase or decrease the observed cross section depending on the relative phase of the nuclear and Coulomb amplitudes. At beam energies above approximately 100~MeV per nucleon and for impact parameters larger than the sum of the projectile and target radii, the nuclear contribution is generally small. Nevertheless, quantitative extractions of electromagnetic response functions require calculations that consistently include both interactions. Glauber theory, dynamical eikonal models, and CDCC calculations provide reliable descriptions of these interference effects over a broad range of energies.

A distinctive consequence of higher-order Coulomb dynamics is the phenomenon of post-acceleration. After breakup, charged fragments continue to experience the Coulomb field of the target and consequently acquire additional kinetic energy while moving away from the interaction region. Because different fragments possess different charge-to-mass ratios, their final momenta differ slightly from those expected from a sudden breakup picture. In proton-rich nuclei, this effect may noticeably shift the relative-energy spectrum and distort fragment momentum distributions. The magnitude of post-acceleration decreases rapidly with increasing beam energy because the interaction time becomes shorter and the projectile trajectory approaches the sudden limit. Time-dependent calculations have demonstrated that post-acceleration is generally small for relativistic Coulomb dissociation experiments but may become appreciable at lower beam energies or for very highly charged targets [\cite{Bertulani2003,Esbensen1996,Typel2001,Yahiro2012,Moro2006,Nunes1996,Capel2004}].

The combined influence of higher multipolarities, nuclear-Coulomb interference, higher-order electromagnetic couplings, and final-state interactions establishes the ultimate theoretical accuracy attainable in Coulomb dissociation analyses. Extensive benchmark calculations using first-order perturbation theory, dynamical Schr\"odinger-equation approaches, CDCC, and dynamical eikonal models have shown that these corrections can usually be controlled at the few-percent level when measurements are restricted to forward scattering angles and large impact parameters. Such theoretical advances have enabled Coulomb dissociation to become one of the most reliable indirect methods to determine radiative capture cross sections, constraining continuum wave functions, and probing the structure of weakly bound nuclei far from stability.

\section{Radioactive beams and Coulomb excitation \label{sec:RIbeam}}

The study of exotic, short-lived nuclei using rare-isotope beams has become a central topic in contemporary nuclear physics~[\cite{NAKA17,CRAW24,Ye2025}]. Because these nuclei lie far from stability, their decay and dissociation reactions provide rich insight into both their intrinsic structure and reaction dynamics. In particular, electromagnetic excitation has attracted sustained interest because it probes collective and single-particle properties in a relatively clean manner. 

Experimentally, such studies often employ heavy, high-$Z$ targets, most commonly 
Pb, at beam energies  
of approximately 70-800 MeV per nucleon~[\cite{BG10,AUMA13,Aumann2020Indirect}]. 
Operating in this intermediate- to high-energy regime helps suppress unwanted contributions from nuclear interactions and reduces the effects 
of complex higher-order breakup processes.

From a theoretical perspective, describing breakup reactions at these energies requires a relativistically consistent treatment of the reaction dynamics. 
In practice, many analyses incorporate relativistic corrections only at the level of kinematics, while retaining non-relativistic dynamics. A more rigorous framework is provided by the virtual photon (equivalent photon) method, which offers a fully relativistic description of projectile excitation induced by the electromagnetic field of the target. 
This approach, however, is limited to electromagnetic processes and does not account for nuclear-induced breakup or multi-step effects. Therefore, to extract physical observables such as reduced transition probabilities  $\beone$  or asymptotic normalization coefficients (ANCs), it is essential to quantify and correct for the effects of nuclear interactions and higher-order contributions to the reaction mechanism. 

In the following sections, we discuss recent developments in Coulomb dissociation experiments 
using rare-isotope beams and relevant theories.
In particular, 
we present representative experiments on neutron halo nuclei and studies of radiative capture reactions relevant to stellar reactions
using the inverse process of Coulomb dissociation, together with theoretical aspects of these experiments.

\section{Halo Nuclei probed by Coulomb dissociation \label{sec:halo}}

\subsection{Neutron Halo Nuclei}

\begin{figure}
\begin{center}
{\includegraphics[width=18.cm]{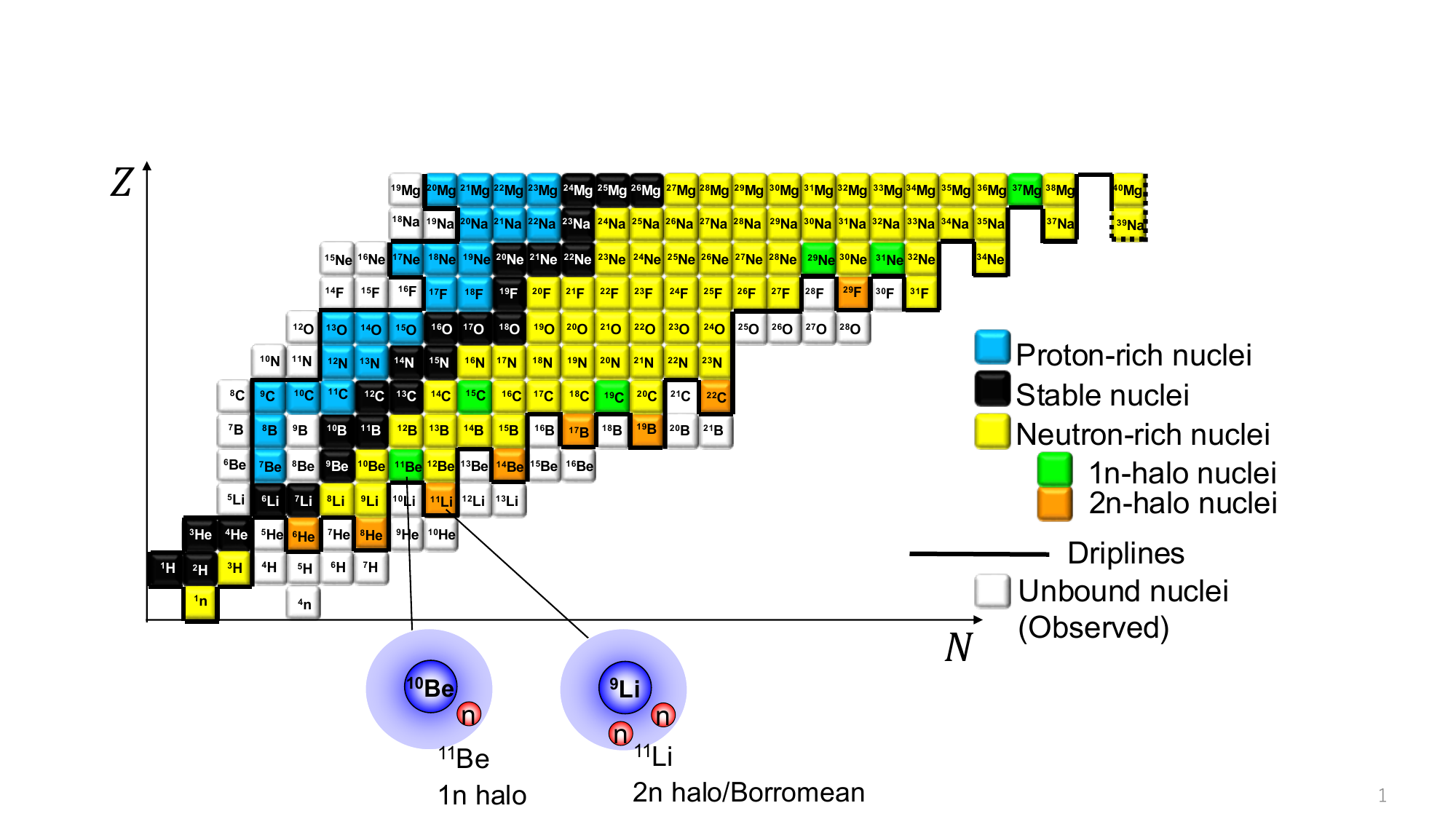}}
\end{center}
\vspace{-0.5cm}
\caption{\label{chart} Neutron halo nuclei in the nuclear chart are shown: one-neutron halo nuclei are marked as green squares
while two-neutron halo nuclei as orange squares. As shown, halo nuclei are observed along the neutron drip line. 
Schematic structures of the one-neutron halo nucleus $^{11}$Be and the two-neutron halo nucleus of $^{11}$Li are depicted.
 }
\end{figure}

Neutron halo nuclei constitute one of the most remarkable manifestations of quantum many-body dynamics in weakly bound systems. They are characterized by one or more neutrons occupying orbitals with very small separation energies, resulting in  spatial wave functions that extend far beyond the range of the nuclear interaction.
Consequently, their matter radii become substantially larger than those predicted by the conventional liquid-drop relation
$
R = r_0 A^{1/3}$,
with $r_0 \simeq 1.2$ fm. 
The first convincing evidence for halo structures came from measurements of the interaction cross sections of light neutron-rich nuclei at relativistic energies, which revealed unexpectedly large matter radii for $^{11}$Li, $^{11,14}$Be, and $^{17}$B~[\cite{Tanihata1985}].
As shown in Fig.~\ref{chart}, $^{11}$Be consists of a $^{10}$Be core surrounded by a single-neutron halo, while $^{11}$Li consists of a  
$^{9}$Li core surrounded by a two-neutron halo.
%This is because the halo is formed by the quantum tunneling for the least bound neutron(s) with very small one or two neutron separation energies below 1 MeV. 
%Currently, the heaviest one-neutron halo nucleus observed is $^{37}$Mg~\cite{Koba14}, while the heaviest two-neutron halo nucleus is $^{29}$F~\cite{}.
%You can find the recent reviews on halo nuclei in Ref.\cite{TANI13}.
%nuclei such as $^{11}$Li and later $^{11}$Be, $^{14}$Be, $^{19}$C, and $^{22}$C [\cite{Tanihata1985,Hansen1987,Tanihata2013}]. 

A hallmark of a neutron halo is the presence of one or two valence neutrons with separation energies typically below 1 MeV.
%Figure \ref{chart} illustrates that the halo nuclei have been observed 
%along the neutron drip line, beyond which the nuclei are no longer bound 
%with respect to one or two neutron emission. 
Accordingly, neutron halo nuclei tend to appear along 
the neutron drip line, as shown in Fig.\ref{chart}. 
%The neutron drip line is defined such that beyond this line the nucleus is no longer bound with respect to one or two neutron emission.
Outside the range of the nuclear potential, the radial wave function decreases exponentially,
\begin{equation}
u_l(r) \propto e^{-\kappa r},
\qquad
\kappa
=
\frac{\sqrt{2\mu S_n}}{\hbar},
\label{eq:halo_tail}
\end{equation}
where $\mu$ is the reduced mass and $\sn$ is the one-neutron separation energy. The characteristic spatial extension of the halo is therefore
$
\xi \sim {1}/{\kappa},
$
which increases rapidly as $S_n\rightarrow 0$. For an $s$-wave halo, the rms radius behaves approximately as
\begin{equation}
\langle r^2\rangle^{1/2}
\propto
\frac{1}{\sqrt{S_n}},
\end{equation}
illustrating that an arbitrarily weakly bound neutron may occupy a region extending tens of femtometers beyond the nuclear core. Larger orbital angular momenta suppress the spatial extension of the neutron wave function because of the centrifugal barrier, making halo formation most favorable for $s$ states and, under suitable conditions, $p$ states [\cite{Jensen2004,TANI13,Hammer2017}].

Halo configurations appear near the neutron drip line  
more frequently than would be expected from a simple shell model. 
In other words, valence neutrons in low-$\ell$
orbitals appear to be favored in neutron-rich nuclei near the drip line.
For instance, the ground state of $^{11}$Be is dominated by the intruder $1s$ orbital rather than the $0p$ orbital expected from the naive shell model.
Similarly, in $^{31}$Ne with $N=21$, the ground state is dominated by the 
$1p_{3/2}$ orbital rather than the $0f_{7/2}$ orbital. Halo formation is 
therefore often closely related to shell evolution.

All known two-neutron halo nuclei are Borromean systems: 
neither of their constituent two-body subsystems, core-$n$ and $n$-$n$, is
bound, whereas the full 
core-$n$-$n$ three-body system is weakly bound. 
This implies that additional binding arises in the three-body system.
Neutron-neutron correlations and continuum coupling are possible mechanisms 
responsible for this Borromean binding. 
Recently, dineutron correlations, corresponding to 
spatially compact $nn$ configurations, have been evidenced 
in $^6$He~[\cite{SUN21}], $^{11}$Li~[\cite{NAKA06,KUBO20}], and $^{19}$B~[\cite{COOK20}], and have become an important topic in studies of two-neutron halo nuclei~[\cite{NAKA26}].

The exotic structural properties of the neutron halo have stimulated extensive theoretical and experimental efforts to understand nuclei near the neutron drip line, where the competition among weak binding, continuum coupling, and many-body correlations produces entirely new forms of nuclear structure~[\cite{Riisager1994,Jensen2004,TANI13}].

The theoretical description of halo nuclei requires an explicit treatment of continuum degrees of freedom and few-body correlations. Modern approaches include continuum 
shell-model methods, the Gamow Shell Model~[\cite{MICH09}], continuum-discretized coupled-channels methods (CDCC)~[\cite{HAGI22,MORO20,MORO25,MATS04}], Faddeev calculations for three-body systems~[\cite{COBI98,COBI98b}], 
and Halo Effective Field Theory 
(Halo EFT)~[\cite{BertulaniHammerVanKolck2002,Hammer2017,HONG22,GOBE23,GOBE24,COST25}]. Halo EFT is particularly attractive because it exploits the separation of scales between the compact core radius, 
$R_{\rm core}$, and the much larger halo radius, $R_{\rm halo}$, allowing observables to be expanded systematically in the small parameter
$
\epsilon
=
{R_{\rm core}}/{R_{\rm halo}}
\ll 1.
$
This framework naturally connects nuclear structure, reactions, and astrophysical capture processes while providing controlled estimates of theoretical 
uncertainties~[\cite{Hammer2017}].

The unique properties of neutron halo nuclei make them ideal laboratories for studying weak binding, quantum tunneling, continuum coupling, and universal phenomena shared with other few-body quantum systems~[\cite{Hammer2017,NAID23,HONG22}]. They also play a central role in modern rare-isotope-beam physics, where measurements of interaction cross sections, momentum distributions following nucleon removal, Coulomb dissociation, transfer reactions, and invariant-mass spectroscopy continue to provide increasingly stringent tests of microscopic nuclear theories [\cite{Jensen2004,Bertulani2004,TANI13,NAKA17}].

\subsection{Coulomb dissociation of halo nuclei and soft $\eone$ excitation}

\begin{figure}
\begin{center}
{\includegraphics[width=\textwidth]{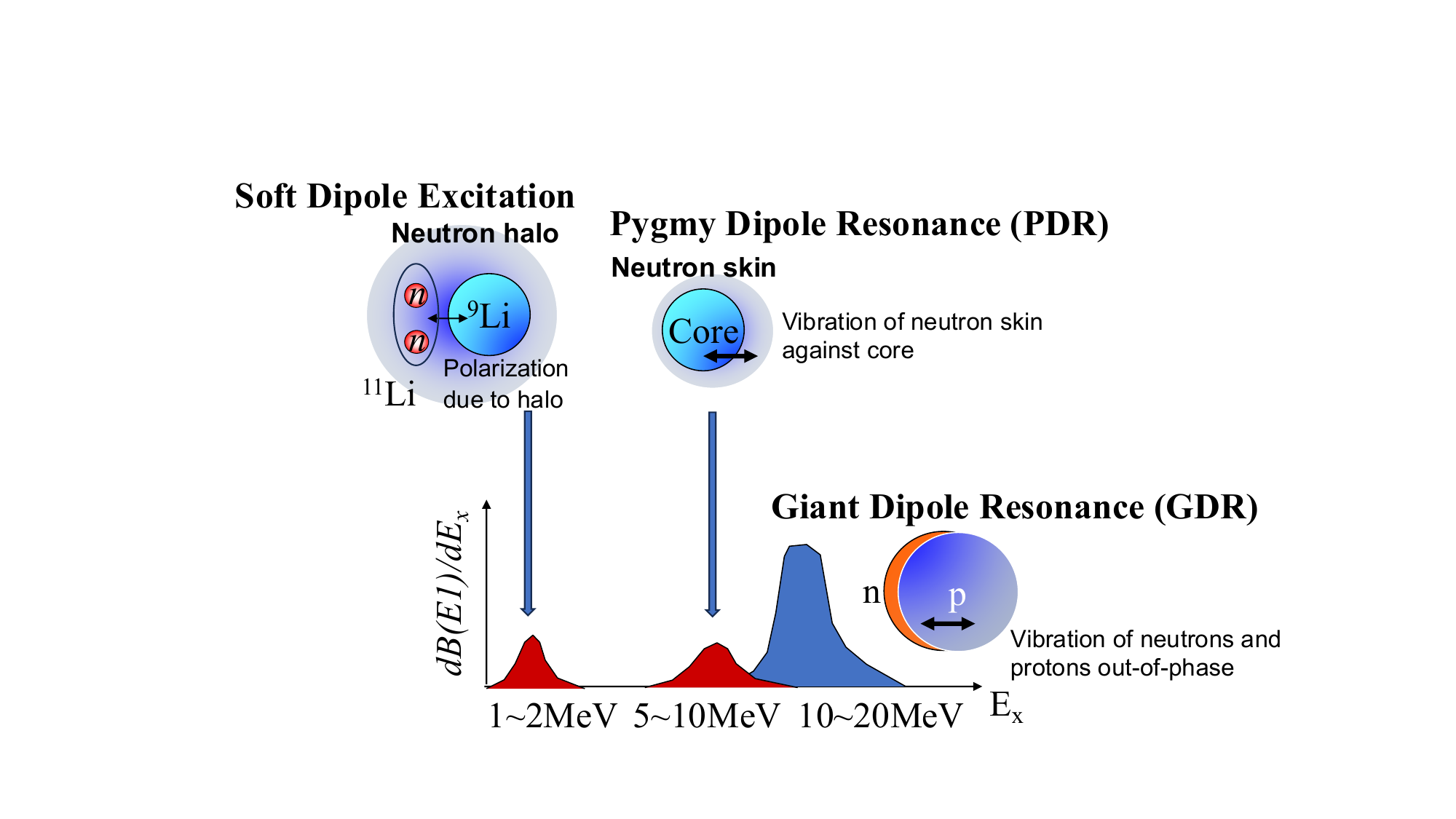}}
\end{center}
\vspace{-0.5cm}
\caption{\label{eoneres} $\eone$ responses of nuclei. For stable nuclei $\eone$ response is exhausted by the giant dipole resonance (GDR) at 10-20 MeV ($\sim 81A^{-1/3}$), with negligible $\eone$ strength at $\ex\sim$ a few MeV, while for neutron halo nuclei,
strong $\eone$ response has been observed at $\ex\sim$~1-2 MeV. In neutron-rich nuclei, we also see pygmy dipole resonance (PDR) around $\ex\sim$ 5-10 MeV.
 }
\end{figure}

The electric dipole response changes systematically from stable 
to neutron-rich nuclei and, eventually, to extremely neutron-rich 
nuclei with neutron halos, as illustrated in Fig.~\ref{eoneres}. 
The $\eone$ response of halo nuclei exhibits a characteristic feature
known as {\it soft $\eone$ excitation}, namely, 
enhanced $\eone$ strengths appearing at excitation energies ($\ex$) of about 1-3 MeV, just above the one-neutron or two-neutron decay threshold 
in one-neutron or two-neutron halo nuclei, respectively. 
This is in sharp contrast to stable nuclei, in which most of the $\eone$ strength is exhausted by
the giant dipole resonance (GDR), corresponding to an out-of-phase vibration of the neutron and proton fluids.
Indeed, it is well known that low-lying $\eone$ strengths are 
substantially hindered in stable nuclei, 
with typical individual strengths of $\beone\sim 10^{-4} - 10^{-6}$ Weisskopf Units (W.u.). 
A strength of $\beone\sim 10^{-2}$~W.u. is already exceptionally large and
generally arises from octupole correlations or other special collective structures~[\cite{ANDR01}]. In contrast, remarkably large $\eone$ strengths 
exceeding 1 W.u below a few MeV have been observed for neutron-halo nuclei.

In neutron-rich nuclei, the pygmy dipole resonance (PDR) may also appear.
The PDR occurs at $\ex\sim$~5-10~MeV,
and is often interpreted as an oscillation of the neutron skin against the core, where the neutron skin represents an excess-neutron 
distribution arising from the neutron-proton asymmetry. 
In unstable nuclei, PDRs have been observed in $^{130,132}$Sn~[\cite{ADRI05}], $^{68}$Ni~[\cite{WIEL09,ROSS13}], and $^{70}$Ni~[\cite{WIEL18}] 
using Coulomb excitation. The PDR strength is correlated with the neutron-skin thickness and with the symmetry energy in the equation of state (EoS) of nuclear matter. For recent studies of PDRs, see the  reviews~[\cite{AUMA13,SAVR13,LANZ23,BRAC19}].

\begin{figure}[h]
 \begin{center}
   \includegraphics[width=10.cm]{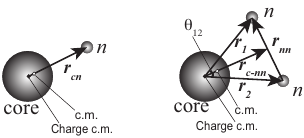}
   \caption{Geometry of a one-neutron halo nucleus (left) and a two-neutron halo nucleus (right). 
   The key geometry in a one-neutron halo nucleus is the relative 
   coordinate vector, $\rcn$. The center of charge (c.m of the core) is dislocated from the center of mass of the entire system,
   which induces the strong soft $\eone$ excitation. 
   A two-neutron halo nucleus is a three-body system, where 
   $\vecrone$, $\vecrtwo$ are the position vectors of respective halo neutrons relative to the c.m. of the core, while 
   $\rcnn$ is that for the c.m. of the two neutrons relative to that of the core. 
   $\theta_{12}$ is the opening angle between $\vecrone$ and $\vecrtwo$.
   The center of charge is dislocated from the c.m. of the whole system when $\langle \theta_{12} \rangle < 180^\circ$.
   }
   \label{fig:halo_coord}
   \end{center}
\end{figure}

The soft $\eone$ excitation shown in Fig.~\ref{eoneres} arises from the
enhancement of low-energy electric dipole ($\eone$) transitions 
associated with the spatially extended neutron distribution in a neutron halo.
Within a simple core-plus-neutron model, 
the reduced transition probability is
\begin{equation}
\frac{d\beone}{dE}
=
\frac{3}{4\pi}
\left(Z_{\rm eff}e\right)^2
\left|
\left<
\psi_f
\left|
rY_{1m}
\right|
\psi_i
\right>
\right|^2
\rho(E),
\label{eq:beone}
\end{equation}
where $Z_{\rm eff}e$ is the effective charge: $Z_{\rm eff}e=Ze/A$ for a one-neutron halo nucleus and  $Z_{\rm eff}e=2Ze/A$ for a two-neutron halo nucleus. 
Here, $r$ denotes the distance from the core to the center of the halo: $r=r_{cn} $ for
a one-neutron halo nucleus and $r_{c-nn}$ for a two-neutron halo nucleus, 
as shown in Fig.~\ref{fig:halo_coord}.
$\rho(E)$ is the continuum level density. 

Because the initial-state wave function extends to very large distances, 
its overlap with low-energy continuum states becomes unusually large, 
producing a pronounced concentration of dipole strength near threshold. 
For one-neutron halo nuclei, we will see more explicit examples below in connection with the so-called direct breakup mechanism.
In this picture, soft $\eone$ excitation does not arise from a specific 
resonance as in GDR and PDR, 
but rather from the strong coupling, or overlap, between the spatially extended halo wave function and the non-resonant continuum. 

Coulomb dissociation experiments exploit this enhancement by measuring the breakup probability in the electromagnetic field of a heavy target. Through the equivalent-photon method, these measurements provide direct information on
the structure of halo nuclei. As shown in Sect.\ref{sec:astro},
radiative neutron-capture reactions of astrophysical importance can also be extracted using detailed balance~[\cite{Baur1986,Bertulani2004,Motobayashi1994,NAKA09}].

%
%As shown in Fig.~\ref{cdfig}, Coulomb excitation (Coulomb dissociation) for halo nuclei due to the weakly bound nature of halo nuclei. As such, the Coulomb dissociation (Coulomb breakup) is a powerful method to observe the $\eone$ response of halo nuclei.

%In this section, we focus on the soft dipole excitation of neutron halo nuclei.
%The $\eone$ response of the halo nuclei can be experimentally studied by its Coulomb dissociation, as the Coulomb excitation leads
%to a decay associated with one- or two-neutron emission due to its weakly binding nature. 

%As detailed below for specific experimental results on soft $\eone$ excitation for halo nuclei,
%the soft $\eone$ excitation is now understood not by a specific resonance as in GDR and PDR, 
%but rather due to the strong coupling (overlap) of the spatially extended wave function (halo wave function)
%with the {\it non-resonant} continuum. This mechanism, called direct breakup, is clearly shown for one-neutron halo nuclei.

Soft $\eone$ excitation can also be interpreted in terms of 
the polarization of neutron-halo nuclei. The center of charge
is displaced from the center of mass (c.m.)  
by $d=r/A$ for a one-neutron halo nucleus and $d=2r/A$ for a two-neutron halo nucleus.
The electric dipole moment is therefore $Ze\times d = Z_{\rm eff}e r$ (see Fig.~\ref{fig:halo_coord}).
The integrated $\eone$ strength associated with 
soft $\eone$ excitation can be expressed
using the non-energy-weighted $\eone$ cluster sum rule~[\cite{ESBE92}], 
\begin{equation}
\beone = \int_{-\infty}^\infty{\frac{d\beone}{d\ex}d\ex}=\frac{3}{4\pi} 
\left(Z_{\rm eff}e\right)^2
%\langle r^2 \rangle=\frac{3}{4\pi} \left(\frac{Ze}{A} \right)^2.
\langle r^2 \rangle.
\label{eq:eonesum1}
\end{equation} 
This expression follows from Eq.(\ref{eq:beone}). 
As shown in the examples below, a halo size of $\sqrt{\langle r^2_{\rm cn} \rangle}\sim5$~fm
corresponds to a soft $\eone$ excitation of approximately 3-4 W.u.
%Some experimental results on such strong soft $\eone$ excitation and the evaluation of the geometry of halo nuclei are shown below.

\subsection{Coulomb dissociation of one-neutron halo nuclei}

One-neutron halo nuclei have been identified, 
as shown in Fig.~\ref{chart}. So far, 
$^{11}$Be,$^{15,19}$C, $^{29,31}$Ne and $^{37}$Mg have been identified as one-neutron halo nuclei.

The one-neutron halo nucleus $^{11}$Be consists 
of a $^{10}$Be core and a single halo neutron,
with one-neutron separation energy of $\sn=$~501.64(25)~keV. 
Although the ground states of $N=7$ isotones normally have spin-parity
$\jpi=1/2^-$, corresponding to occupation of the $0p_{1/2}$ single-particle orbital, the ground state $^{11}$Be has $\jpi=1/2^+$, arising from the intruder $1s_{1/2}$ orbital. Thus, $^{11}$Be
is a representative $s$-wave one-neutron halo nucleus and has been studied extensively both experimentally and theoretically.
%The Coulomb dissociation experiments have been conducted three times and one of them is detailed below.

Another notable $s-$wave halo nucleus is $^{19}$C, 
which consists of an $^{18}$C core and a single halo neutron.
Coulomb dissociation of $^{19}$C was studied in Ref.~[\cite{NAKA99}], 
establishing its $s$-wave halo nature with
the one-neutron separation energy of about 0.5 MeV, close to that of $^{11}$Be. 

\newcommand{\ngamma}{(n,\gamma)}
\newcommand{\nalpha}{(n,\alpha)}
\newcommand{\betami}{(\beta^-)}

The heavier one-neutron halo nuclei $^{29,31}$Ne and $^{37}$Mg
all exhibit $p$-wave halo configurations. For example, the representative 
$p$-wave halo nucleus $^{31}$Ne consists of a $^{30}$Ne core coupled to a $p$-wave single halo neutron. The
spin-parity of $^{31}$Ne is therefore 3/2$^-$. 
This contrasts with the naive shell-model picture for $N=21$ isotones, 
in which the ground-state spin-parity is expected to be 7/2$^-$ because
of occupation of the $0f_{7/2}$
orbital. This indicates that the $N=20$ magicity is lost at $^{31}$Ne. Indeed, neutron-rich Ne isotopes near $N=20$ 
belong to the island of inversion~[\cite{WARB90,OTSU20,NOWA21}], and $^{31}$Ne is no exception.

This $p$-wave halo nature of these nuclei has been investigated 
through inclusive Coulomb-dissociation measurements for $^{29}$Ne~[\cite{KOBA16}], 
$^{31}$Ne~[\cite{NAKA09b,NAKA14}] and $^{37}$Mg~[\cite{KOBA14}].
In an inclusive Coulomb dissociation measurement, the neutron emitted in the
decay of the excited halo nucleus is not detected, and therefore the $\beone$
distribution cannot be reconstructed. However, inclusive measurements 
have the advantage of providing sufficient statistics even with low-intensity rare-isotope beams. 
%This is particularly important for very exotic nuclei and 
%during the early stages of operation of a new facility.
%Such inclusive Coulomb-dissociation experiments were performed
%during the early years of RIBF at RIKEN.
Nevertheless, as discussed below, 
useful spectroscopic information could still be extracted, as demonstrated
by the Coulomb-dissociation measurements on $^{31}$Ne~[\cite{NAKA09b,NAKA14}].

\subsubsection{Coulomb dissociation of $^{11}$Be}

Coulomb dissociation measurements of $^{11}$Be have been performed
at 72 MeV/nucleon~[\cite{NAKA94}], 69 MeV/nucleon at RIKEN~[\cite{FUKU04}], and 520 MeV/nucleon at GSI~[\cite{PALI03}].
In these experiments,
the breakup of $^{11}$Be on a Pb target into 
$^{10}$Be and a neutron was measured.
In the coincidence exclusive measurements, the invariant-mass
spectrum, or equivalently the breakup cross section as a function of the two-body relative energy, can be reconstructed. 
In the higher-energy experiment~[\cite{PALI03}], deexcitation $\gamma$ rays
were also measured, and the decay probability to $^{10}$Be(2$^+$) was found to become more significant at higher excitation energies because of the larger number of virtual photons (see Fig.~\ref{vpnfig}). 

Overall agreement among the $\beone$ distributions obtained in 
these experiments is found, except for their absolute amplitudes: 
the $\beone$ strength reported in Ref.~[\cite{PALI03}] is 
approximately 20\% smaller than those obtained in the two RIKEN
experiments~[\cite{NAKA94,FUKU04}]. 
This discrepancy is currently interpreted as arising from nuclear distortion, higher order coupling effects, 
and Coulomb-nuclear interference terms~[\cite{MORO20}], which are not 
included in the equivalent photon method.

\begin{figure}[h]
 \begin{center}
   \includegraphics[width=10.cm]{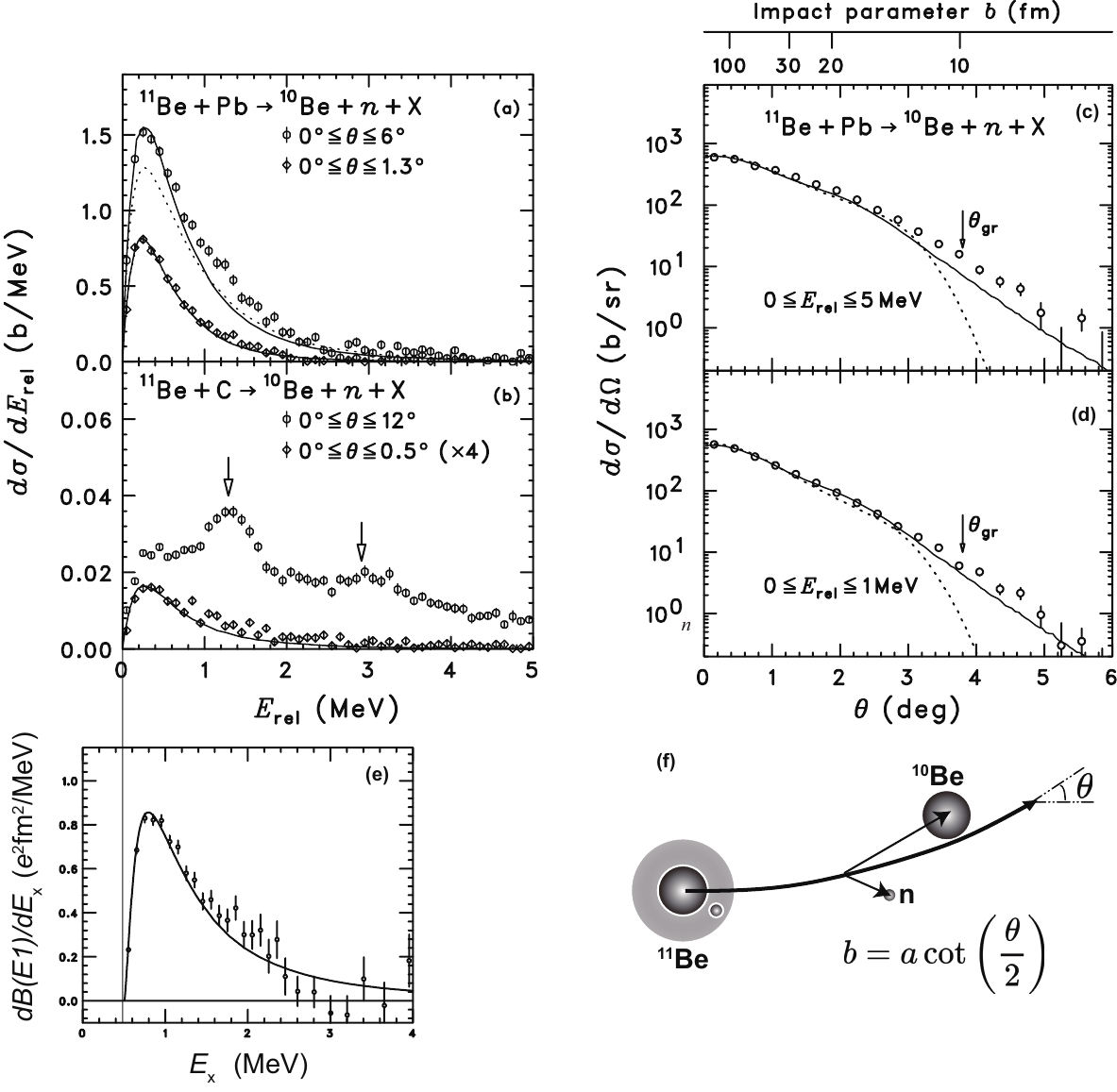}
   \caption{(a) Energy-differential dissociation cross sections for $^{11}$Be+Pb 
   into $^{10}$Be+$n$ for the angular ranges $0\leq\theta\leq 6^\circ$ 
   and $0\leq\theta\leq 1.3^\circ$, where the scattering angle $\theta$ 
   is defined as the opening
   angle between the momentum vectors of the incoming $^{11}$Be and 
   the outgoing $^{11}$Be$^*$ in the c.m. 
   system of $^{11}$Be+Pb. The definition of the scattering angle is schematically illustrated in (f).
   (b) Energy-differential dissociation cross sections for $^{11}$Be+C 
   into $^{10}$Be+$n$ for the angular ranges $0\leq\theta\leq 12^\circ$ 
   and $0\leq\theta\leq 0.5^\circ$. The scattering angle is defined in
   the same way as for $^{11}$Be+Pb, but 
   for the c.m. system of $^{11}$Be+C. 
   The solid curves show calculations of the direct-breakup mechanism for soft $\eone$ excitation in 
   $^{11}$Be using the ECIS code, whereas the dotted curves show direct-breakup 
   calculations using the equivalent photon method. The ECIS and equivalent-photon calculations agree for the forward-angle selected data.
   (c), (d) Scattering-angle distributions for $^{11}$Be+Pb
   in the selected $\erel$ regions: (c) $0\leq\erel\leq$~5~MeV, and 
   (d) $0\leq\erel\leq$~1~MeV. The impact parameter $b$ is related to $\theta$ as $b=a \cot(\theta/2)$.
   As shown in (c) and (d), the selection of $\theta\leq 1.3^\circ$ 
   lies well below the grazing angle
   $\theta_{\rm gr}$. 
   (e) The $\beone$ distribution obtained using the forward-angle data 
   ($\theta\leq 1.3^\circ$)
   in (a). Figs.~(a)-(d) are reproduced from Ref.~[\cite{FUKU04}], whereas Fig.~(e) is reproduced from~[\cite{NAKA2012}]. 
   }
   \label{fig:coulbu_11be}
   \end{center}
\end{figure}

Here, we focus on the high-statistics dissociation data of $^{11}$Be+Pb at 69 MeV/nucleon and $^{11}$Be+C at 68 MeV/nucleon 
obtained at RIKEN~[\cite{FUKU04}].
Figure~\ref{fig:coulbu_11be} shows the results of this experiment: (a) energy differential breakup cross sections 
for $^{11}$Be + Pb for two selected angular ranges, 
(b) the corresponding results for $^{11}$Be + C, 
(c) angular distributions for 
$^{11}$Be + Pb with $\erel\leq 5$~MeV, and (d) those with $\erel\leq$ 1 MeV.
%Here, we explain how the $\beone$ spectrum in Fig.~\ref{fig:coulbu_11be}(e) is extracted from these.

An important issue in Coulomb dissociation experiments 
is how to separate the nuclear-breakup component, which is 
present even for a high-$Z$ target
such as Pb. Two methods have been developed to extract the 
Coulomb dissociation component from the total dissociation cross section:
(1) subtraction of the nuclear-breakup component using carbon-target data, 
and (2) extraction of 
the Coulomb dissociation component by selecting forward-angle scattering events. Ref.~[\cite{FUKU04}] was the first experiment to adopt
Method (2), whereas Refs~[\cite{NAKA94,PALI03}] used Method (1).

Method (1) assumes that the carbon-target data arise solely
from nuclear breakup. The nuclear-breakup component
for a halo nucleus on a Pb target is then evaluated by scaling the carbon-target data according to 
\begin{equation}
\frac{d \sigma}{d E_{\mathrm{rel}}}=\frac{d \sigma_{\mathrm{Pb}}}{d E_{\mathrm{rel}}}-\Gamma \frac{d \sigma_{\mathrm{C}}}{d E_{\mathrm{rel}}},
\label{eq:meth1}
\end{equation}
where $d\sigma/d\erel$, $d\sigma_{\mathrm{Pb}}/d\erel$,
and $d\sigma_{\mathrm{C}}/d\erel$ denote the energy-differential Coulomb-dissociation cross section, the breakup cross section on a Pb target,
and that on a C target, respectively.
The scaling factor $\Gamma$ can be evaluated using 
simple geometrical arguments~[\cite{NAKA94}], eikonal calculations~[\cite{PALI03}], 
or CDCC calculations~[\cite{COOK20}]. 
In the Coulomb dissociation of $^8$He discussed below, 
additional measurements with medium heavy targets improved the precision
of this evaluation. 

The Coulomb dissociation experiment of Ref.~[\cite{FUKU04}] made full use 
of the measured scattering angle $\theta$. It was determined from the incident momentum vector $\vecbeam$ and the sum of the outgoing momentum vectors,
$\vecfrag+\vecneut$, which corresponds to the momentum vector of the outgoing $^{11}$Be center-of-mass system. Namely, 
the opening angle between $\vecbeam$ and $\vecfrag+\vecneut$, evaluated
in the center of mass of $^{11}$Be + Pb, provides the scattering angle of $^{11}$Be. 
As illustrated in Fig.~\ref{fig:coulbu_11be}(f), the Coulomb-excitation process approximately follows a Rutherford trajectory, and
$\theta$ can therefore be regarded as the Rutherford scattering angle to a good approximation. 
The impact parameter $b$ is then related to $\theta$ by
\begin{equation}
b=a \cot \left(\frac{\theta}{2}\right),
\end{equation}
where $a$ is half the distance of closest approach in Rutherford scattering. 

The validity of the angular-cut method is demonstrated in Figs.~\ref{fig:coulbu_11be}~(c) and \ref{fig:coulbu_11be}~(d). 
The angular distributions are
well reproduced by the ECIS code (DWBA, Distorted-Wave Born Approximation) and 
by the equivalent photon method below the grazing angle $\theta_{\rm gr}$. 
The angular cut $\theta\leq 1.3^\circ$, which is sufficiently smaller than the grazing angle $\theta_{\rm gr}=3.8^\circ$, 
corresponds to an impact-parameter cut of $b\geq 30$~fm under the 
assumption of Rutherford scattering.

%Figure~\ref{fig:coulbu_11be}(f) shows schematically the method of extacting $\theta$ and $b$, and (c)(d) are the angular distributions 
%for $\erel\leq$~5MeV and $\erel\leq$~1MeV. $\theta_{\rm gr}$(=3.8$^\circ$) represents the grazing angle of $^{11}$Be + Pb at 69 MeV/nucleon. 

The energy-differential cross sections in Fig.~\ref{fig:coulbu_11be}(a) are presented for $\theta\leq 6^\circ$ and 
$\theta\leq 1.3^\circ$, and are compared with 
calculations based on the direct-breakup mechanism.
In the equivalent photon method,
the energy differential cross section is expressed as
\begin{equation}
\dsderel=\frac{16\pi^3}{9\hbar c}\neone\dbderel.
\end{equation}
Substituting the direct-breakup expression for 
$\dbde$, 
\begin{equation}
\dsderel= \frac{16\pi^3}{9\hbar c}\neone\mid \langle{\bf q} \mid \frac{Ze}{A}rY_{1m} \mid \psi_i\rangle \mid^2,
\label{eq:directbu}
\end{equation}
where $\mid\psi_i\rangle$ represents the initial state dominated by the 
neutron-halo component, $rY_{1m}$ is the $\eone$ operator, with $r$ being
the distance between the core and neutron ($r\equiv r_{cn}$ in Fig.\ref{fig:halo_coord}~(left)), and
$\langle{\bf q}\mid$ represents the non-resonant 
continuum state with relative momentum ${\bf q}$. 
The quantity $\dbde$ in this expression corresponds to Eq.~(\ref{eq:beone})
for non-resonant final states.
$^{11}$Be dissociates into $^{10}$Be+n without forming a resonance.
We also use Eqs.~(\ref{sigmagamma}) and (\ref{sig}), with $\ex=\egam$.
In the plane-wave limit, the transition matrix element essentially maps the spatial distribution of the halo neutron onto the continuum momentum distribution. Consequently, the extended tail of the halo wave function strongly enhances the $\eone$ response at small relative energies.
%The direct breakup model can be viewed as a Fourier Bessel transform of the halo wave function. Hence, the large amplitude of the halo wave function at large distances produces a large $\beone$ strength at low relative energies. 
%For low-lying continuum states, 
%$\erel=(\hbar q)^2/(2\mu)$.
The direct breakup model reproduces the energy distribution for 
the data with $\theta\leq 1.3^\circ$, implying that the forward-angle events 
are predominantly due to Coulomb dissociation and that the nuclear-breakup component is negligible.
The $\beone$ distribution extracted from the $\theta\leq 1.3^\circ$ data 
in Fig.\ref{fig:coulbu_11be}(e) is therefore well reproduced by 
the direct-breakup calculation shown by the solid curve. 
No subtraction such as that in Eq.~(\ref{eq:meth1})
is required in this method.

The success of the direct-breakup description of soft $\eone$ excitation 
demonstrates that Coulomb dissociation can serve as a useful spectroscopic tool for constraining the halo wave function.
 The ground-state wave function of $^{11}$Be can be written as
\begin{equation}
 \psi_i =  \alpha\mid\shalfzero\rangle+ \beta\mid\dhalftwo\rangle+ ...,
% \psi_i =  \alpha\mid\shalfzero\rangle+ \beta\mid\dhalftwo\rangle+ ...,
 \label{eq:wfbe11}
\end{equation}
where $\alpha^2$ and $\beta^2$ represent the corresponding spectroscopic factors. 
The magnitude of the $\beone$ distribution and 
the corresponding Coulomb-dissociation cross sections can 
be used to extract $\alpha^2$ because  
soft $\eone$ excitation is primarily sensitive to the halo component
represented by the first term of Eq.(\ref{eq:wfbe11}). This yields 
$\alpha^2=0.72\pm 0.04$~[\cite{FUKU04}].

The peak position in the spectrum 
is sensitive to the one-neutron separation energy $\sn$ 
and the orbital angular momentum $\ell$ of the halo state.
For an $s$-wave single neutron halo, the $\beone$ spectrum peaks 
at $\erel=(3/5)\sn$ in the plane-wave approximation, 
as this approximation provides the analytical form for the 
$\beone$ distribution~[\cite{BERTULANI1988615,NAGA05,NAKA94}], 
\begin{equation}
\dbderel\propto\frac{\sqrt{\sn}\erel^{3/2}}{(\erel+\sn)^4}.
\end{equation}

The halo size can be estimated using Eq.(\ref{eq:eonesum1}).
The integrated strength up to $\ex=$~4 MeV is $\beone=1.05(6) e^2$fm$^2$, corresponding to 3.29(19) W.u, and gives $\sqrt{\langle r_{cn}^2\rangle}=5.77(16)$~fm. 

Remarkably, even for the carbon target data shown 
in Fig.~\ref{fig:coulbu_11be}~(b), the forward-angle events with 
$\theta\leq 0.5^\circ$ are
described by direct Coulomb breakup, as shown by the solid curve. 
For the wider angular range, however, two resonance peaks appear at 
$\erel=$1.28 MeV ($\ex=$1.78~MeV), and $\erel=$~2.91 MeV ($\ex=$3.41~MeV), which
are produced by nuclear interactions.

For the $^{11}$Be+Pb data with $\theta\leq 6^\circ$, 
the data show overall agreement with the direct-breakup calculations with ECIS (DWBA, first-order perturbation) and the equivalent-photon method, 
although some deviation due to nuclear breakup remains. 
We examine the effectiveness of Method (1), the subtraction method.
Most of the deviation is removed using 
Eq.~(\ref{eq:meth1}) with $\Gamma=2.1$, although some deviation due to 
nuclear breakup remains  
at $\erel\sim$~1-2 MeV, as shown in Fig.~7 of Ref.~[\cite{FUKU04}].
The remaining discrepancy may arise from higher-order effects or 
Coulomb-nuclear interference. The use of Method (1) with a constant 
$\Gamma$ should therefore be treated with some caution. 
Further experimental and theoretical studies will be needed 
for future high-precision Coulomb dissociation measurements.

%\subsubsection{$^{19}$C}
\subsubsection{$^{31}$Ne}

We now move from $^{11}$Be ($N=7$)  to $^{31}$Ne ($N=21$). 
As discussed above, $^{11}$Be is exotic in two respects: 
it has a neutron halo, and the conventional $N=8$ shell closure is
broken.
The latter property is associated with strong deformation of 
the $^{10}$Be core.
$^{31}$Ne is analogous to $^{11}$Be in that it has a neutron halo 
and also exhibits the breaking of the conventional 
$N=20$ shell closure. It has been shown that 
not only the $N=20$ but also the $N=28$ shell gap is reduced 
in this region, and that 
the $^{30}$Ne core is well deformed.
Such exotic properties of $^{31}$Ne 
have been revealed through Coulomb dissociation measurements~[\cite{NAKA09b,NAKA14}].

The dissociation of $^{31}$Ne on Pb and C targets was studied 
during the early stage of RIBF operation~[\cite{NAKA09,NAKA14}]. 
RIBF started operation in 2007~[\cite{MOTO12}]. 
Because the secondary-beam intensity was limited to 
about
five particles per second at that time, 
inclusive Coulomb-dissociation measurements 
were performed for the reaction $^{31}$Ne + Pb$\rightarrow ^{30}$Ne$+X$.
In these inclusive measurements, the emitted neutron was not detected,
and only the one-neutron 
removal cross section was obtained.
In the equivalent photon method, this corresponds to measuring the integrated cross section
\begin{equation}
\sigmac = 
%\sigmaonenc = 
%\int_{\sn}^\infty{\dsde d\ex} = 
%\int_{\sn}^{E_u}{
\int_{\sn}^{\infty}{
\frac{16\pi^3}{9\hbar c}\neone\dbdex d\ex}.
\label{eqbeone}
\end{equation}
Because the virtual-photon spectrum decreases rapidly with excitation 
energy (\fig{vpnfig}), even the integrated Coulomb-dissociation cross section can provide a clear signature of soft $\eone$ excitation.
If the two-neutron emission channel, which opens at higher excitation energies, is also taken into account, the one-neutron removal cross section is dominated by the soft $\eone$ excitation. 
Indeed, the first inclusive measurement of the one-neutron removal cross section 
of $^{31}$Ne on a Pb target at 240 MeV/nucleon already showed evidence for 
the one-neutron halo structure of $^{31}$Ne~[\cite{NAKA09b}].

Further information can be obtained by
combining the Coulomb-dissociation cross section with the nuclear-breakup cross section measured using a carbon target, as was done in the second experiment~[\cite{NAKA14}].
Additional $\gamma$-ray coincidence measurements made it possible
to extract partial cross sections leading to the excited and ground states of $^{30}$Ne.
In $^{30}$Ne, the first 2$^+$ state at $\ex=$800 keV is a 
possible final state, and its decay to the ground state can be identified through the emitted $\gamma$ ray.
By subtracting the contribution associated with the excited-core component, one can extract the one-neutron removal cross sections leading directly 
to the $^{30}$Ne ground state for Coulomb dissociation, $\sigmaeonegs$, 
and for the carbon target, $\sigmacgs$.
The measured values were
$\sigmaeonegs=$448(108)~mb, 
and $\sigmacgs=$33(15)~mb.

\begin{figure}[h]
 \centering
   \includegraphics[width=8cm]{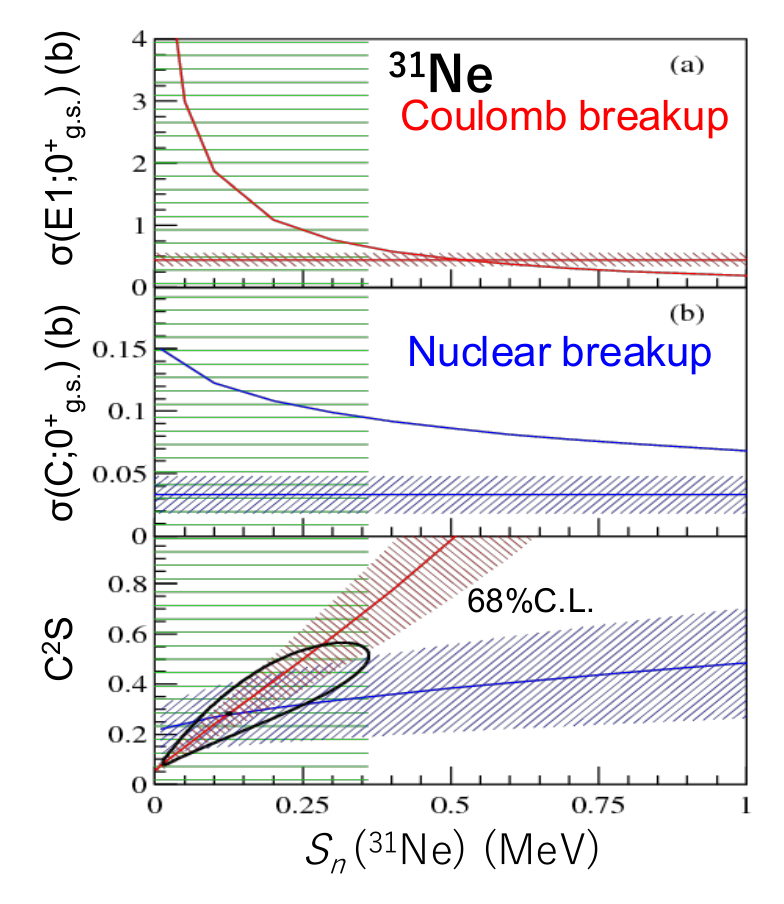}
   \caption{Top: The experimental Coulomb-breakup cross section feeding directly the $^{30}$Ne ground state, 
   $\sigmaeonegs=$448(108)~mb, is compared with the calculated cross section
   based on the direct breakup model for the $\mid\pconf\rangle$ configuration
   of the $^{31}$Ne ground state with $\jpi=3/2^-$, shown as a function of $\sn$.
   The green band indicates the experimental upper limit on $\sn$~[\cite{GAUD12}].
   Middle: The experimental one-neutron removal cross section on a carbon target feeding the $^{30}$Ne ground state, $\sigmacgs=$33(15)~mb, is compared with the eikonal-model calculation for the same configuration.
   Bottom: Constraints on the spectroscopic factor $\cs$ as a function of 
   $\sn$ obtained from Coulomb and nuclear breakup. The red and blue bands correspond to the constraints from Coulomb and nuclear breakup, respectively, and the black contour indicates the combined 68\% confidence region.
   }
   \label{fig:coulbu_31ne}
  % \end{center}
\end{figure}

Figure \ref{fig:coulbu_31ne} compares the experimental values of
$\sigmaeonegs$ and $\sigmacgs$ with theoretical calculations for
Coulomb and nuclear breakup, shown by the red and blue solid curves 
in panels (a) and (b), respectively.
The calculated cross sections are shown as functions of $\sn$,
since the one-neutron separation energy 
of $^{31}$Ne is only poorly known.
A direct 
 mass measurement provided only an upper limit, 
 $\sn\leq$~360~keV, shown by the green band in 
 Fig.~\ref{fig:coulbu_31ne}~[\cite{GAUD12}]. 
The Coulomb-dissociation calculation is based on the direct-breakup mechanism and the equivalent-photon method of $\eqn{eq:directbu}$, integrated as in Eq.~(\ref{eqbeone}). For nuclear breakup, i.e., one-neutron removal on a
carbon target, the eikonal model was applied. 

These calculations were performed assuming a $^{31}$Ne ground state with spin-parity $3/2^-$, and the spectroscopic factor $\cs$=1  
for the configuration $\mid\pconf\rangle$.
The ratio of the experimental cross section to the calculated one 
therefore provides an estimate of $\cs$ 
as a function of $\sn$. Because Coulomb and nuclear-breakup cross sections have different dependences on $\sn$, the allowed values of 
$\cs$ and $\sn$ can be constrained simultaneously, as shown by the 
black contour in the bottom panel of Fig.~\ref{fig:coulbu_31ne}.
This analysis gives $\sn=0.15^{+0.16}_{-0.10}$~MeV and $\cs=0.32^{+0.21}_{-0.17}$ for the  
$\mid\pconf\rangle$ configuration.

This result demonstrates that $^{31}$Ne has a $p$-wave one-neutron halo.
The large Coulomb dissociation 
cross section is incompatible with 
the $\mid\fconf\rangle$ configuration 
expected from the naive shell-model picture.
The experimental result is therefore 
consistent with $^{31}$Ne lying in the island of inversion. 

In $^{31}$Ne, the near degeneracy of the $0f_{7/2}$ and $1p_{3/2}$ orbitals in the spherical limit favors the development of deformation
through the Jahn-Teller effect, 
as discussed by Hamamoto~[\cite{HAMA10}].
The contribution of low-$\ell$ orbitals, such as $1p$, 
becomes increasingly important  
as $\sn$ approaches zero. This mechanism may be referred to as a {\it deformation-driven halo}. 
Hamamoto suggested that the valence neutron in $^{31}$Ne 
may occupy either the Nilsson orbital [330]1/2 
for moderate prolate deformation, $0.22 \leq\beta\leq 0.30$, 
or [321]3/2 for stronger prolate deformation, $0.40 \leq\beta\leq 0.59$~[\cite{HAMA10}].

The structure of $^{31}$Ne should be further clarified by 
exclusive Coulomb-dissociation measurements with higher statistics. Such a measurement has recently been performed 
at RIBF, RIKEN, and is expected to clarify how deformation and 
halo formation are related.

Soft $\eone$ excitation 
has also been observed in the neighboring 
nuclei $^{29}$Ne~[\cite{KOBA16}] and $^{37}$Mg~[\cite{KOBA14}], 
and has likewise been attributed to $p$-wave halo structures. 
For a recent review of these measurements, see~[\cite{NAKA23}].
%Recently, the $N=21$ fluorine isotope, $^{30}$F was observed as a 
%$p$-wave unbound state, which favors the picture of $^{31}$F~\cite{KAHL24} as $p$-wave two neutron halo, as in $^{29}$F~\cite{BAGC20}.
This suggests that halos may also occur in heavier nuclei
when low-$\ell$, particularly $s-$ or $p$-wave, components
emerge as a result of shell evolution near the neutron drip line. 
The search for halos in heavier drip-line nuclei therefore 
continues, and 
Coulomb dissociation will remain a powerful tool for this purpose.

\subsection{Coulomb dissociation of two-neutron halo nuclei}

Two-neutron halo structures have been 
identified in $^{6}$He, $^{11}$Li, $^{14}$Be, $^{17,19}$B, $^{22}$C, and $^{29}$F.
Interestingly, within each isotopic chain, the nucleus at the neutron drip line appears to have a high probability of exhibiting a two-neutron halo structure. 
The neutron drip line has so far been experimentally established only up to the Ne isotopes~[\cite{AHN19}], 
and among the known drip-line nuclei, clear exceptions to this tendency are 
$^3$H and $^{24}$O. Other neutron-drip-line nuclei, such as $^{31}$F and 
$^{34}$Ne, whose spectroscopy has not yet been fully established, may also exhibit two-neutron halo structures.
$^8$He may be regarded as a mixture of two-neutron-halo and four-neutron-skin configurations. As discussed below, soft $\eone$ excitation associated with the 
$^6$He+$2n$ three-body configuration has been observed~[\cite{DUER26}], whereas the two- and four-neutron removal cross sections~[\cite{TANI92}] and the charge radius~[\cite{MUEL07}] favor a four-neutron-skin picture.

%$^8$He may be regarded a mixture of two-neutron halo and four-neutron skin configurations. 
%As shown below, soft $\eone$ excitation due to 
%the $^6$He+$2n$ three-body system has been observed~[ 
%while the $2n$- and $4n$-removal cross sections~, 
%as well as charge radii~[\cite{MUEL07}], favor the four-neutron skin.

The configurations of the two halo neutrons vary among these drip-line nuclei.
In $^6$He and $^8$He, the valence neutrons predominantly occupy the $0p_{3/2}$ orbital, and thus $p$-wave configurations dominate. The two-neutron halo in 
$^{11}$Li is a mixture of $(1s)^2$ and $(0p)^2$ configurations. A recent quasi-free-scattering experiment using the $^{11}$Li$(p,pn)$ 
reaction showed that the two-neutron configurations in 
$^{11}$Li contain $(1s)^2$ and $(0p)^2$ components of 35(4)\% and 59(1)\%,
respectively~[\cite{KUBO20}].
For $^{14}$Be, the $^{14}$Be$(p,pn)$ experiment favors an $(1s)^2$ contribution of approximately 20\% and a $(0p)^2$ contribution of approximately 60\% [\cite{CORS19}], similar to those in $^{11}$Li. 
%The valence neutrons in $^6$He and $^8$He lie mainly in $0p_{3/2}$, so the $p$-wave halo is dominant. The two-neutron halo in $^{11}$Li is a mixture of $(1s)^2$ and $(0p)^2$. The recent quasi-free scattering of $^{11}$Li$(p,pn)$ showed
%two-neutron configurations in $^{11}$Li 
%re $(1s)^2$:35(4)\% and $(0p)^2$: 59(1)\%.
%For $^{14}$Be, the $^{14}$Be$(p,pn)$ experiment favors the $(1s)^2$ contribution 
%of $\sim$20\% and the $(0p)^2$ contribution of $\sim$60\%~[\cite{CORS19}], 
%close to $^{11}$Li. 
The situation for $^{17}$B remains controversial. Interaction cross section measurements suggest a large $(1s)^2$ component, whereas the recent $^{17}$B$(p,pn)$ experiment shows only a 9(2)\% $(1s)^2$ component~[\cite{YANG21}]. The $^{19}$B nucleus exhibits a two-neutron halo with a significant $s$-wave component~[\cite{COOK20}]. 
Studies of $^{22}$C are still limited, 
although $s$-wave neutrons are expected to play an important role in halo formation. $^{29}$F has recently been indicated to have a $p$-wave halo 
associated with the $1p_{3/2}$ component, suggesting that it belongs to the island of inversion~[\cite{BAGC20}]. A $p$-wave halo is also expected in $^{31}$F~[\cite{KAHL24}].
Coulomb dissociation is an important method for investigating the 
configurations and correlations of the two halo neutrons.

The first Coulomb dissociation on a halo nucleus was performed for $^{11}$Li
on Pb at 0.79 GeV/nucleon by Kobayashi et al. at LBNL in 1989~[\cite{KOBA89}].
Then, in the 1990s, three exclusive Coulomb dissociation experiments on $^{11}$Li were performed, in which the momenta of $^{9}$Li and the two neutrons were measured 
in coincidence~[\cite{ieki1992,shimoura1995coulomb,zinser1997invariant}].
Later, a Coulomb-dissociation experiment with higher statistics
and improved sensitivity near $\erel\sim$~0 MeV was performed 
at RIKEN at 70 MeV/nucleon~[\cite{NAKA06}]. Exclusive Coulomb dissociation
has also been measured for $^6$He~[\cite{AUMANN1996321,SUN21}], $^{14}$Be~[\cite{LABI01}], 
$^{19}$B~[\cite{COOK20}], and $^{8}$He~[\cite{DUER26}].
Here, we focus on the Coulomb dissociation experiments on $^{11}$Li and $^{8}$He. 

%As shown in the nuclear chart~\ref{chart}, neutron drip line shapes the staggering structure.
%For instance, for helium isotopes, $^{5}$He, $^{7}$He are unbound while $^{6}$He, and $^{8}$He are bound.
%For lithium, $^{10}$Li is unbound while $^{11}$Li is bound. 

\subsubsection{Coulomb dissociation of $^{11}$Li}

\begin{figure}[h]
 \begin{center}
   \includegraphics[width=7.cm]{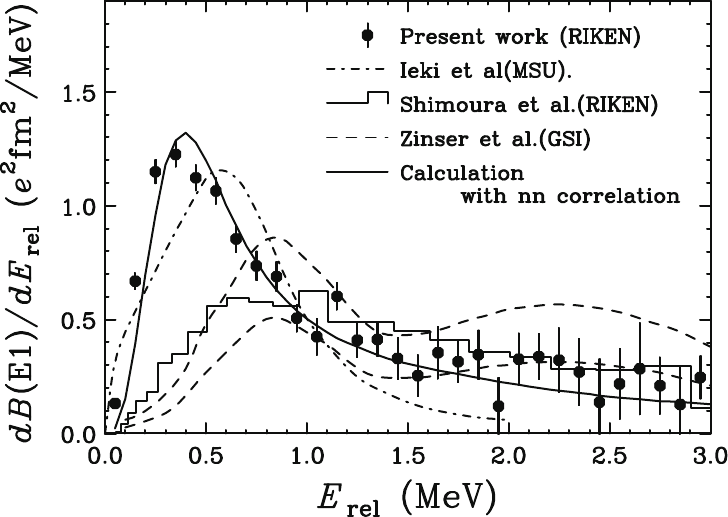}
   \caption{The observed $\dbde$ distribution for $^{11}$Li obtained from Coulomb dissociation on a Pb target at 70 MeV/nucleon~[\cite{NAKA06}]. The solid line shows the three-body calculation~[\cite{ESBE92}].
   The earlier results~[\cite{ieki1992,shimoura1995coulomb,zinser1997invariant}] are also shown; these measurements had lower statistics and reduced sensitivity 
   either near $\erel=$~0 or at $\erel>$ 2 MeV.
   The figure is reproduced from Ref.~[\cite{NAKA06}].}
   \label{fig:coulbu_11li}
   \end{center}
\end{figure}

The $\beone$ distribution for $^{11}$Li as a function of the three-body relative energy ($^9$Li+$n$+$n$), obtained from Coulomb dissociation
of $^{11}$Li on Pb at 70 MeV/nucleon, 
is shown in Fig.~\ref{fig:coulbu_11li}~[\cite{NAKA06}].
It shows a sharp peak at $\erel\sim 0.3$ MeV. 
The spectrum is compared with the three-body calculation (solid line)~[\cite{ESBE92}], 
which is in good agreement with the data.
The earlier measurements had lower statistics and less sensitivity 
either near $\erel=0$~MeV~[\cite{zinser1997invariant,shimoura1995coulomb}] or 
at higher $\erel$~[\cite{ieki1992}], and are therefore not discussed here.

Coulomb dissociation of two-neutron halo nuclei provides information on the geometry of the three-body system (see Fig.~\ref{fig:halo_coord}).
The non-energy weighted cluster sum rule given in Eq.~(\ref{eq:eonesum1}) can be applied to a two-neutron halo nucleus as 
\begin{equation}
\beone = \int_{-\infty}^\infty{\frac{d\beone}{d\ex}}d\ex
          =\frac{3}{4\pi} \left(\frac{Ze}{A} \right)^2
            \langle (\vecrone+\vecrtwo)^2 \rangle \\
            = \frac{3}{4\pi} \left(\frac{Ze}{A} \right)^2
            \langle r_1^2+r_2^2+ 2\vecrone\cdot \vecrtwo \rangle \\
            = \frac{3}{\pi}\left(\frac{Ze}{A} \right)^2 \langle r_{c-nn}^2 \rangle,
            \label{eq:eonesum2}
\end{equation}
where $\vecrone$, $\vecrtwo$, and $\bf r_{c-nn}$ denote 
the position vectors defined in 
Fig.~\ref{fig:halo_coord}~(right). 
The sum rule allows us to determine the rms distance from the core to the c.m. of the two halo neutrons, while $\vecrone\cdot\vecrtwo$
contains information on the $nn$ spatial correlation.

In Ref.~[\cite{NAKA06}], we assumed that the $\eone$ response beyond $\erel=3$~MeV follows the three-body model, which
reproduces the measured spectrum (solid curve), and extracted an 
integrated strength of $\beone$=1.78(22) e$^2$fm$^2$. From this value, 
$\sqrt{\langle r^2_{c-nn}\rangle}$=5.01(32)~fm was extracted. 
Combining this result with the non-correlated $\eone$ strength evaluated using the three-body model,
we obtained the opening angle $\langle\theta_{12}\rangle={48^\circ}^{+14}_{-18}$.

Alternatively, $\langle\theta_{12}\rangle$ can be extracted
using  
the following relation~[\cite{HASA07,NAKA26,BERT07}]:
\begin{equation}
\langle r_m^2 \rangle=\frac{A_c}{A}\langle r_m^2 \rangle_c
+ \frac{2A_c}{A^2}\langle r^2_{c-nn}\rangle + \frac{1}{2A}\langle r^2_{nn} \rangle,
\label{eq:rnn}
\end{equation}
where $\langle r_m^2 \rangle$ and $\langle r_m^2 \rangle_c$ represent the mean-square matter radii of
the two-neutron halo nucleus with mass number $A$ 
and its core with mass number $A_c=A-2$, respectively.
Using this relation,  
$\langle\theta_{12}\rangle=56.2^{+17.8}_{-13.1}$ degrees was obtained~[\cite{HASA07}], consistent with the value given above.
The opening angle smaller than 90$^\circ$ indicates 
a strong spatial $nn$ correlation, referred to as a {\it dineutron correlation}.
The mixture of the opposite-parity 
$(1s)^2$ and $(0p)^2$ configurations plays a crucial role
in generating this correlation. See the recent review 
on dineutron clustering~[\cite{NAKA26}].

%\subsubsection{Coulomb dissociation of $^{19}$B}

\subsubsection{Coulomb dissociation of $^{8}$He}

$^8$He is the most neutron-rich bound nucleus, with $N/Z=3$, on the neutron drip line. 
However, its two-neutron separation energy, $\snn=2.1250(1)$ MeV, 
is larger than that of $^6$He, $\snn=0.975457(5)$~MeV, 
indicating the extra stability of $^8$He associated with 
the $N=6$ sub-shell closure.
$^8$He is considered to consist of a $^4$He core 
surrounded by a neutron skin with four neutrons, which is denser 
than a neutron halo. 
This picture is experimentally supported~[\cite{TANI92}] by the following relations involving the interaction cross sections of $^{4,6,8}$He on carbon at 800 MeV/nucleon and
the two- and four-neutron removal cross sections of $^8$He:
\begin{eqnarray}
\sigma_I(^8{\rm He})-\sigma_I(^4{\rm He}) &\approx& \sigma_{-2n}(^8{\rm He})+ \sigma_{-4n}(^8{\rm He}). \\
\sigma_I(^8{\rm He})-\sigma_I(^{6}{\rm He}) &\ll& \sigma_{-2n}(^8{\rm He}),
\end{eqnarray}
where $\sigma_I(^AZ)$ is the interaction cross section for the $^AZ$ isotope, while
$\sigma_{-2n}(^8{\rm He})$ and $\sigma_{-4n}(^8{\rm He})$ represent the 
two- and four-neutron removal cross sections of $^8$He, respectively.
These relations show that $^8$He is more closely associated
with 
the $^{4}$He+$4n$ configuration than with the $^{6}$He+$2n$ configuration.
The charge radius of $^8$He measured by isotope-shift spectroscopy is smaller than that
of $^6$He, indicating that, while $^6$He has a two-neutron halo with a dineutron correlation, the four valence neutrons in $^8$He 
have a more spherically symmetric distribution~[\cite{MUEL07}].

Coulomb dissociation of $^8$He was previously measured 
at GSI at 227 MeV/nucleon, suggesting a possible 
soft-$\eone$-excitation peak at $\ex\sim$~4 MeV 
($\erel\sim$ 2~MeV)~[\cite{MEIS03}]. This experiment measured only the $^6$He+2n decay channel and had limited statistics.
Recently, Coulomb dissociation of $^8$He was measured at RIBF
at $\sim$185 MeV/nucleon for both the $^6$He+$2n$ and $^4$He+$4n$ 
decay channels.
It should be noted that the coincidence detection of multiple neutrons
at intermediate energies has long been challenging because of neutron cross talk, and the unambiguous detection of more than two neutrons has been particularly difficult~[\cite{NAKA16,KOND20}].
Before the $^8$He Coulomb-dissociation experiment, 
the only case in which the momenta of four neutrons at intermediate energies had been measured unambiguously in coincidence was the first observation of 
$^{28}$O decaying into $^{24}$O+$4n$~[\cite{KOND23}].

\begin{figure}[h]
 \begin{center}
   \includegraphics[width=12.cm]{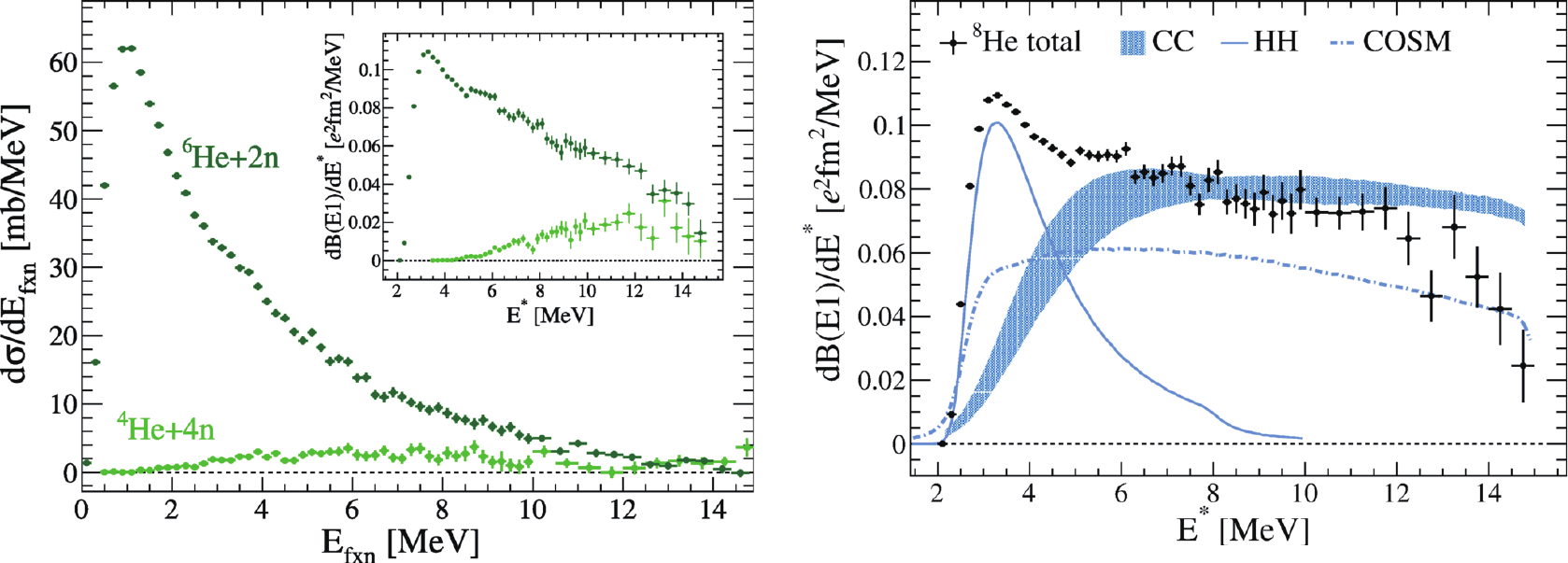}
   \caption{(Left) Coulomb-dissociation cross sections of $^8$He on Pb at 185 MeV/nucleon for the three-body $^6$He+$2n$ decay channel (dark-green points) and the 
   five-body $^4$He+$4n$ decay channel (light-green points). The cross sections are plotted as functions of $E_{fxn}$, which represents either the three-body relative energy, $E_{f2n}$,  or the five-body relative energy, $E_{f4n}$, depending on the decay channel. (Right) The extracted $\beone$ strength distribution as a function of the excitation energy $E^*$.
   The two- and four-neutron decay thresholds 
   are $E^*=$2.125~MeV and 3.1~MeV, respectively. The data are compared with 
   coupled-cluster (CC) calculations~[\cite{BONA22}], 
   a three-body-model calculation~[\cite{GRIG09}],
   and a cluster-orbital shell-model (COSM) calculation~[\cite{MYO22}]. The figures are reproduced from Ref.~[\cite{DUER26}].
   }
   \label{fig:coulbu_8he}
   \end{center}
\end{figure}

The Coulomb-dissociation cross sections as functions of the three-body relative energy for the $^6$He+$2n$ channel and the five-body relative energy for the $^4$He+$4n$ channel are shown in Fig.~\ref{fig:coulbu_8he} (left).
As shown, even above the $4n$ decay threshold ($E_{fxn}=0.97$~MeV when
expressed in terms of the three-body relative energy), the $^6$He+$2n$ channel is by far dominant.
The $^{6}$He+$2n$ cross section is strongly enhanced just above 
the two-neutron decay threshold, a characteristic feature of soft $\eone$ excitation. The dominance of the $^6$He+$2n$ channel 
even above the $^4$He+$4n$ threshold may support the
interpretation that the soft $\eone$ excitation is associated
with the $^6$He+2n configuration.

%%%%% kokomade 2026.8.7

As shown in Fig.~\ref{fig:coulbu_8he}~(right), 
the extracted $\dbdest$, where $E^*$ represents the excitation energy of $^8$He, exhibits a narrow peak at $E^*\sim$ 3MeV ($E_{f2n}\sim$~1 MeV) and a broader enhancement at $E^*\sim$ 5-12 MeV. 
Interestingly, the first peak is reproduced 
by the three-body-model calculation~[\cite{GRIG09}], whereas the broader enhancement is reproduced by the ab initio coupled-cluster calculation~[\cite{BONA22}].
This may indicate that the latter calculation does not include sufficient correlations
to reproduce the component associated with the $^6$He + $2n$ halo structure. 
Thus, the first peak may be attributed 
to soft $\eone$ excitation, as in other two-neutron halo nuclei such as $^6$He~[\cite{AUMANN1996321,SUN21}], $^{11}$Li~[\cite{NAKA06}], and $^{19}$B~[\cite{COOK20}], whereas the enhancement
at higher energies may be more closely related to the $^4$He+$4n$ structure.

On the other hand, a five-body cluster-model calculation, in which the unbound states are described using the complex scaling method and the motion 
of the four valence neutrons around the $^4$He core is treated within the cluster orbital shell model (COSM), shows an overall
enhancement of the low-lying $\eone$ strength at $E^*\sim 3-12$~MeV, although the calculated strength is smaller than the experimental one. This calculation indicates the importance of dipole excitation associated with the 
$^7$He(g.s.)+$n$ configuration. We note that sequential decay through the $^7$He ground state is observed experimentally. However, because the experimental $\beone$ distribution is not reproduced by the COSM calculation, 
it is difficult to draw a firm conclusion about the role of the $^7$He +$n$ component.
Further experimental and theoretical investigations of the ground and excited states of $^8$He, as well as of the dynamics of its breakup reactions,
are therefore needed.

\section{Coulomb dissociation for astrophysical applications\label{sec:astro}}

Radiative capture reactions play a central role in nuclear astrophysics. In these processes, two nuclei (or a nucleon and a nucleus) combine to form a bound system while emitting a photon,
$
a + b \rightarrow c + \gamma .
$
Because the electromagnetic interaction removes excess energy and angular momentum, radiative capture provides a primary mechanism for element synthesis in stellar environments. Many key reactions in hydrogen burning, helium burning, and explosive nucleosynthesis proceed through this channel. Direct measurements of radiative capture cross sections at astrophysical energies are extremely challenging. For charged-particle reactions at stellar energies (typically in the keV range), cross sections are strongly suppressed by the Coulomb barrier [\cite{Baur1986}].
This difficulty has motivated the development of indirect techniques, among which the Coulomb dissociation method is one of the most powerful.

Radiative capture reactions often involve charged particles, such as
%$
%(p,\gamma) \quad \text{and} \quad (\alpha,\gamma),
%$
$(p,\gamma)$ and $(\alpha,\gamma)$,
as well as neutron capture reactions,
$
(n,\gamma),
$
which are essential for the $s$- and $r$-processes of nucleosynthesis.
To factor out the dominant Coulomb suppression in charged-particle reactions, one introduces the astrophysical $S$-factor,
\begin{equation}
\sigma(E) = \frac{S(E)}{E} \exp(-2\pi\eta),
\end{equation}
where $\eta$ is the Sommerfeld parameter. The function $S(E)$ varies slowly with energy and contains the relevant nuclear structure information. Accurate knowledge of $S(E)$ is crucial for modeling stellar evolution, solar neutrino production, and explosive astrophysical scenarios.

The Coulomb dissociation method provides an indirect way to determine radiative capture cross sections by studying the inverse reaction. Instead of measuring  [\cite{Baur1986}]
$
a + b \rightarrow c + \gamma,
$
one investigates
$
c + T \rightarrow a + b + T,
$
where the projectile $c$ interacts with the strong electromagnetic field of a heavy target $T$ (e.g., %$^{208}\mathrm{Pb}$
Pb) (see Fig. \ref{cdfig}). Using Eq. \eqref{dsig}  and the principle of detailed balance, the photo-absorption cross section for radiative capture reactions, $\gamma + c \rightarrow a + b$, can be related to the radiative capture cross section of astrophysical interest, $a + b \rightarrow c + \gamma$. Thus, measurements of Coulomb dissociation at intermediate or high energies allow the extraction of low-energy capture information.

The Coulomb dissociation technique offers several advantages: (a) Breakup cross sections at intermediate energies (50-250 MeV/nucleon) are much larger than direct capture cross sections at astrophysical energies. (b) The method is especially effective for loosely bound and halo nuclei. (c) Electric dipole ($E1$) transitions, which often dominate astrophysical capture processes, are strongly enhanced.
The method has been successfully applied to reactions such as
$^{7}\mathrm{Be}(p,\gamma)^{8}\mathrm{B}$ [\cite{EBS05}],
relevant to solar neutrino production, as well as to neutron and proton capture reactions important in explosive nucleosynthesis.
Accurate extraction of capture rates requires careful theoretical treatment. Important aspects include: (i) Separation of electromagnetic and nuclear breakup contributions, (ii) inclusion of higher-order and multistep effects, (iii) relativistic corrections at intermediate beam energies,
and (iv) contributions from higher multipolarities. Reaction models such as first-order perturbation theory and continuum-discretized coupled-channels (CDCC) calculations are commonly employed. Reliable modeling is essential for deducing the reduced transition probabilities $B(E\lambda)$, asymptotic normalization coefficients (ANCs) and astrophysical $S$-factors.  When combined with robust reaction theory and precise experiments, it yields crucial nuclear data to understand stellar evolution and the origin of the elements.

\subsection{Coulomb dissociation of $^{8}$B}

\begin{figure}[h]
\begin{center}
\includegraphics[width=8.cm]{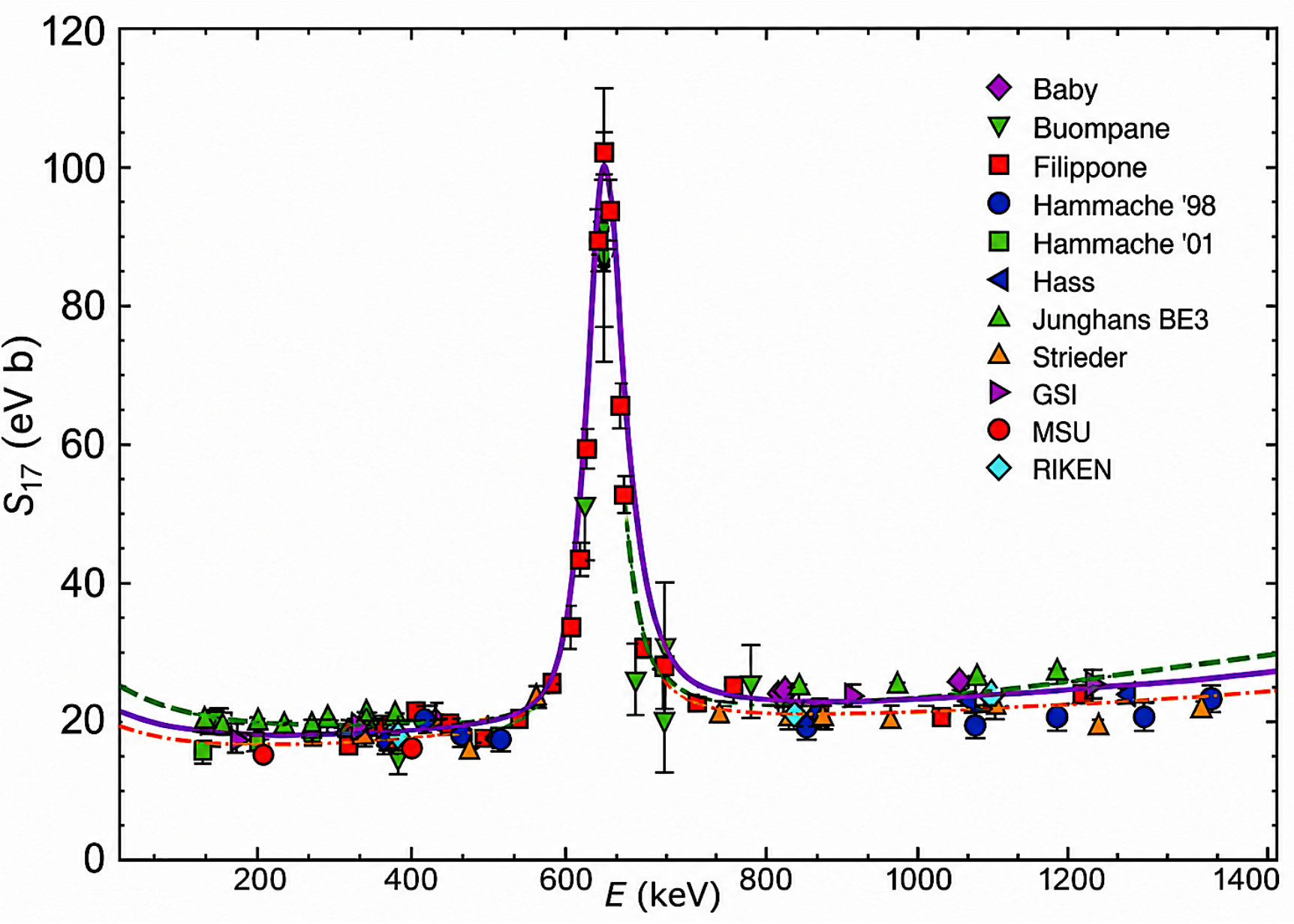}
\caption{Measured astrophysical $S$ factor for the
$^{7}\mathrm{Be}(p,\gamma)^{8}\mathrm{B}$ capture reaction together
with Bayesian fits obtained using halo effective field theory (Halo
EFT). The analysis includes only data with center-of-mass energies
$E_{\mathrm{c.m.}}\leq 1250$~keV and employs the Halo EFT
formulation of [\cite{Higa2022}], which
accounts for the excited-state configuration of the
$^{7}\mathrm{Be}$ core. The figure illustrates the progressive
improvement of the EFT description from leading order (LO) through
next-to-leading order (NLO) to next-to-next-to-leading order (NNLO).
The calculations include the excited $^7$Be core and the $\mone$ contribution around the 1$^+$ resonance. For more details and references, see [\cite{Acharya2025}], from which the figure is reproduced.} 
   \label{fig:s17}
\end{center}
\end{figure}

The radiative-capture reaction $^7\mathrm{Be}(p,\gamma)^8\mathrm{B}$ is a crucial branch of the solar proton-proton chain because the subsequent $\beta^+$ decay of $^8\mathrm{B}$ produces the highest-energy component of the solar-neutrino spectrum [\cite{Adelberger2011}]. At the low center-of-mass energies relevant to the solar interior, the capture cross section is extremely small. The reaction proceeds predominantly through nonresonant electric-dipole ($\eone$) capture from incoming $s$- and $d$-wave proton states into the $J^\pi=2^+$ ground state of $^8\mathrm{B}$. Because the proton separation energy is only
$
S_p(^8\mathrm{B})
\simeq
137~\mathrm{keV},
$
the capture is highly peripheral and is governed mainly by the asymptotic part of the $^7\mathrm{Be}+p$ overlap function. A $J^\pi=1^+$ resonance at $E_{\mathrm{c.m.}}\simeq 0.63~\mathrm{MeV}$ produces an important magnetic-dipole ($M1$) contribution at intermediate energies, although it has little influence on the extrapolation to $E=0$. The Solar Fusion III evaluation [\cite{Acharya2025}] recommends
\begin{equation}
S_{17}(0)
=
20.5\pm0.7~\mathrm{eV,b},
\end{equation}
based on a combined analysis of direct-capture and Coulomb-dissociation measurements using halo effective field theory. For a current data-and-theory summary, the most useful figure is the $S_{17}(E)$ compilation in Solar Fusion III [\cite{Acharya2025}] redrawn 
in Fig.~\ref{fig:s17}. It combines the principal direct $^7\mathrm{Be}(p,\gamma)^8\mathrm{B}$ data sets and the accepted Coulomb-dissociation results with the leading-order, next-to-leading-order, and next-to-next-to-leading-order halo-EFT calculations [\cite{Higa2022}].  The GSI spectra and extracted $S_{17}(E)$ values reported by Sch\"umann \textit{et al.} remain among the clearest examples [\cite{Schuemann2003,Schuemann2006}]. 

 Major measurements using the Coulomb dissociation of $^8\mathrm{B}$ were performed at RIKEN, GSI, Michigan State University, and Notre Dame [\cite{Motobayashi1994,Iwasa1996,Kikuchi1998,Iwasa1999,Guimaraes2000,Davids2001PRL,Schuemann2003,Schuemann2006}] over beam energies extending from near the Coulomb barrier to approximately $254~\mathrm{MeV}$/nucleon. The relativistic GSI measurements, performed with restrictive forward-angle cuts, showed that the breakup is dominated by $E1$ excitation and yielded values of $S_{17}(0)$ consistent with direct-capture measurements [\cite{Schuemann2003,Schuemann2006}].
The extraction of $S_{17}(E)$ from Coulomb-breakup data nevertheless requires a detailed reaction model. In addition to the dominant first-order $E1$ contribution, the measured cross section can contain electric-quadrupole $E2$ excitation, $E1$--$E2$ interference, nuclear breakup, Coulomb-nuclear interference, and higher-order electromagnetic effects [\cite{Bertulani1994,EsbensenBertsch1995,Bertulani1996,BertulaniGai1998,Bertulani2005RCDCC,OgataBertulani2010,Zhang2015,Higa2022,Zhang2018}]. Schematically, the breakup amplitude can be expressed as
\begin{equation}
\mathcal{M}
=
\mathcal{M}*{E1}
+
\mathcal{M}*{E2}
+
\mathcal{M}*{\mathrm{nucl}}
+
\mathcal{M}*{\mathrm{higher}},
\end{equation}
so that
$
|\mathcal{M}|^2
\neq
|\mathcal{M}*{E1}|^2
+
|\mathcal{M}*{E2}|^2
+
|\mathcal{M}_{\mathrm{nucl}}|^2,
$
because interference terms may be important. These corrections depend strongly on the beam energy, the maximum projectile scattering angle, the minimum impact parameter, and the experimental acceptance. At relativistic energies and very forward angles, first-order equivalent-photon calculations are generally reliable, whereas experiments at lower energies often require continuum-discretized coupled-channels, dynamical-eikonal, or time-dependent calculations. Earlier disagreements concerning the magnitude of the $E2$ component were one reason why Coulomb-dissociation results were not included in the recommended Solar Fusion II average. The Solar Fusion III analysis reconsidered the RIKEN, GSI, and MSU measurements, assigned additional systematic uncertainties associated with the assumed $E2$ strength, and found that the direct and indirect determinations of $S_{17}(0)$ are mutually consistent [\cite{Acharya2025,Adelberger2011}].

Modern calculations of $^7\mathrm{Be}(p,\gamma)^8\mathrm{B}$ employ potential models, microscopic cluster models, $R$-matrix theory, ab initio methods, and halo effective field theory
[\cite{BertulaniHammerVanKolck2002,NavratilBertulaniCaurier2006,NavratilRothQuaglioni2011,NavratilQuaglioni2012,Zhang2015,Higa2022,Zhang2018}]. In a peripheral description, the low-energy capture amplitude is primarily determined by the asymptotic-normalization coefficients $C_1$ and $C_2$ of the two possible channel-spin $S=1$ and $S=2$ configurations. The zero-energy astrophysical factor therefore has the approximate dependence
$
S_{17}(0)
\propto
C_1^2+C_2^2,
$
although short-distance operators, proton-$^7\mathrm{Be}$ scattering parameters, and coupling to the first excited state of $^7\mathrm{Be}$ also contribute. Halo EFT is particularly useful because the small binding momentum,
$
\gamma
=
{\sqrt{2\mu S_p}}/{\hbar},
$
provides a natural low-momentum scale that is well separated from the scale associated with the internal structure of the $^7\mathrm{Be}$ core. The theory can consequently be organized as a controlled expansion in a parameter of the form
$
\epsilon
\sim
{Q}/{\Lambda},
$
where $Q$ represents the low-energy momentum scale and $\Lambda$ is the breakdown scale. Current next-to-next-to-leading-order analyses simultaneously describe direct-capture and Coulomb-breakup data, include the $1^+$ resonant $M1$ contribution, and propagate experimental normalization and theoretical truncation uncertainties. Further progress would benefit from new high-statistics measurements below $E\simeq300~\mathrm{keV}$ and from exclusive breakup experiments providing relative-energy, angular, and longitudinal-momentum distributions that can separately constrain the $E1$, $E2$, nuclear, and higher-order components [\cite{BertulaniHammerVanKolck2002,Zhang2015,Zhang2018,Higa2022,Acharya2025}].

\subsection{Coulomb dissociation of $^{15}$C}

$^{15}$C may be regarded as a moderate $s$-wave halo nucleus,
in the sense that it has a relatively large one-neutron separation energy
of $\sn=$~1.2181(8)~MeV compared
with those of $^{11}$Be and $^{19}$C, which are about 0.5 MeV.
Its Coulomb dissociation is
important for determining the radiative-capture cross section
of $^{14}$C$(n,\gamma)$$^{15}$C. This reaction is relevant
to the neutron-induced CNO cycle,
$^{14}$C$\ngamma$
$^{15}$C$\betami$
$^{15}$N$\ngamma$
$^{16}$N$\betami$
$^{16}$O$\ngamma$
$^{17}$O$\nalpha$
$^{14}$C
, which is one of the possible stellar reactions in the asymptotic giant branch (AGB) stars~[\cite{WIES99}]. Among the reactions in this chain, $^{14}$C$\ngamma^{15}$C is the slowest and thus 
determines the overall reaction rate of the cycle.

The cross section of the radiative neutron-capture reaction $^{14}$C$(n,\gamma)$ 
can be extracted from the inverse reaction, Coulomb dissociation of $^{15}$C, using the principle of detailed balance.
\begin{equation}
\sigma_{n \gamma}\left(E_{\text {c.m. }}\right)=\frac{2 I_A+1}{2 I_{A-1}+1} \frac{E_\gamma{ }^2}{2 \mu c^2 E_{\text {c.m. }}} \sigma_{\gamma n}\left(E_\gamma\right),
\end{equation}
where $\sigma_{n \gamma}$ is the radiative neutron-capture cross section, while
$\sigma_{\gamma, n}(=\sigma_\gamma^{\eone})$ is the photoabsorption cross section appearing in Eqs.(\ref{dsig})(\ref{sigmagamma})(\ref{sig}). 
$I_A$ and $I_{A-1}$ are the spins of
$^{15}$C and $^{14}$C, respectively. 
%--- kokomade 2026.8.8 12:24
Owing to the kinematical factor and the large number of virtual photons,
the Coulomb-breakup cross section exceeds 
$\sim 100$~mb/MeV, whereas $\sigma_{n\gamma}$ is only of the order of 10$\mu$b. This demonstrates the advantage of the Coulomb dissociation over the direct neutron-capture measurements.

\begin{figure}[h]
 \begin{center}
   \includegraphics[width=7.cm]{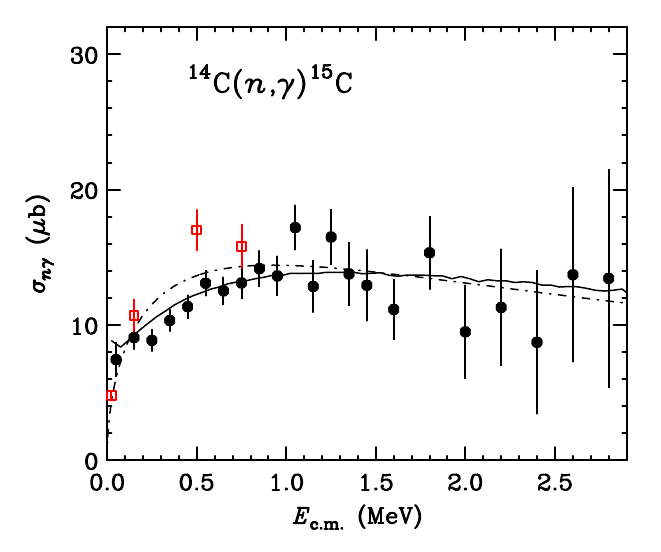}
   \caption{The black solid circles show 
   the $^{14}$C$(n,\gamma)$$^{15}$C cross sections
   as a function of the neutron-$^{14}$C c.m. energy, extracted from the Coulomb dissociation of $^{15}$C, while the red squares show the results of the direct neutron-capture measurement~[\cite{REIF08}]. The solid curve shows the direct neutron-capture calculation folded with the experimental resolution, whereas the dot-dashed curve shows the same calculation before folding. 
   The figure is reproduced
   from Ref.~[\cite{NAKA09}].
   }
   \label{fig:ngam14c}
   \end{center}
\end{figure}

A Coulomb-dissociation measurement at 68 MeV/nucleon~[\cite{NAKA09}] showed 
that soft $\eone$ excitation occurs in $^{15}$C. 
Figure~\ref{fig:ngam14c} shows the $\sigma_{n\gamma}$ cross sections 
extracted from this Coulomb dissociation
(black solid circles).
The red squares show the results of the direct neutron-capture measurement ~[\cite{REIF08}], which are consistent with the Coulomb-dissociation results. The curves show direct neutron-capture calculations, corresponding
to the inverse of the direct breakup process.
The result is also consistent with that obtained from a higher-energy 
Coulomb-dissociation experiment at 605 MeV/nucleon~[\cite{DATT03}]. 

Since the $\eone$ transition in Coulomb dissociation proceeds
from an $s$-wave bound state to the $p$-wave continuum,
the inverse radiative-capture reaction is dominated by 
a $p$-to-$s$ transition, i.e., $p$-wave direct capture, which is 
unusual for a stellar neutron-capture reaction. 
Neutron capture is usually dominated by $s$-wave neutrons, for which 
the cross section is enhanced at low energies according to
$\sigma_{n\gamma}\sim 1/v\sim1/\sqrt{E}$. 
In contrast, the $p$-wave neutron capture cross section behaves
approximately as $\sigma_{n\gamma}\sim\sqrt{E}$, and therefore 
becomes more important at higher stellar temperatures.

This provides one of the clearest examples in which 
Coulomb dissociation can be used to determine the cross section of the inverse $(n,\gamma)$ reaction.
% kokomade 13:48 2026.8.8
The $^{A-1}Z\ngamma ^AZ$ reaction may also populate 
bound excited states of $^AZ$, which introduces an additional complication. For $^{15}$C, there is only one bound excited state, whose contribution 
to $\sigma_{n\gamma}$ is negligible. 
For heavier and more strongly bound nuclei, however, this method has
some limitations.
Coulomb dissociation is particularly powerful for weakly bound 
nuclei, as exemplified by both $^{15}$C and $^8$B.

%\section{Coulomb dissociation for assessing stellar nuclear reaction rates}
%\subsection{Coulomb dissociation of $^8$B}

\section{Summary and Perspectives \label{sec:summary}}

Since the pioneering Coulomb-dissociation experiments on $^{11}$Li
in 1989 at LBNL~[\cite{KOBA89}] and on $^8$B in 1994 at RIKEN~[\cite{Motobayashi1994,Iwasa1996}], 
Coulomb dissociation of rare isotopes has become a powerful experimental tool. In this method, exotic nuclei are excited and dissociated through
the absorption of quasi-real photons generated by a high-Z target at relativistic energies.
The equivalent photon method, in which the Coulomb-excitation cross section is factorized in terms of the virtual-photon number and the reduced transition probability, has enabled us to access key transition strengths
that provide structural information on exotic nuclei, as well as 
reaction rates relevant to stellar processes. 
Because the process is predominantly induced by the electromagnetic interaction, its theoretical treatment is relatively straightforward.
Nevertheless, in some cases, higher-order effects and nuclear-breakup effects should be treated more carefully. Theoretical developments, such as the  continuum-discretized coupled-channels (CDCC) method, have helped to better understand the experimental results.

Accordingly, Coulomb dissociation has become one of the standard experimental methods using rare-isotope beams 
for investigating the microscopic structure of neutron-halo nuclei and for determining radiative-capture reaction rates relevant to stellar processes.
Currently, most Coulomb-dissociation experiments employ 
kinematically complete measurements.

Coulomb-dissociation experiments on halo nuclei have 
revealed a characteristic 
$\eone$ response known as soft $\eone$ excitation, in which $\beone$ strengths exceeding one W.u. appear just above the 
one- or two-neutron-decay threshold 
at $\ex\sim$ 1-3 MeV. 
The integrated $\eone$ strength provides a direct measure
of the halo size, characterized by the rms distance $\sqrt{\langle r^2 \rangle}$.

The $\eone$ response of one-neutron halo nuclei is well understood:
soft $\eone$ excitation arises predominantly through
the direct-breakup mechanism without forming a resonance.
%, and the low-lying $\beone$ strength is attributed to the strong overlap between the extended halo wave function multiplied by $\eone$ operator and low-energy waves of continuum.
For one-neutron halo nuclei, the low-energy $\eone$ response is mainly determined by the spatially extended halo component of the ground-state wave function and its coupling to the nonresonant continuum. Consequently, the experimentally-extracted $\dbde$ distribution provides constraints on the neutron separation energy, orbital configuration, and spectroscopic strength of the halo component.
In addition, the rms core-halo distance, $\sqrt{\langle r_{cn}^2 \rangle}$, has been extracted to be about 5-6 fm using 
the non-energy-weighted $\eone$ sum rule. 

For two-neutron halo nuclei, the $\eone$ response requires more elaborate theoretical descriptions, such as three-body models. 
%For $^6$He, $^{11}$Li, $^{19}$B, the three-body models reproduced well the experimental data.
Nevertheless, the integrated $\eone$ strength is related through the 
non-energy-weighted cluster sum rule to the mean-square distance 
 $\langle r_{c-nn}^2 \rangle$ between the core and the c.m. of the two halo neutrons. This information can also be used
to evaluate the neutron-neutron correlation.
The spatial dineutron correlation has been studied for $^{6}$He~[\cite{SUN21}], $^{11}$Li~[\cite{NAKA06}], and $^{19}$B~[\cite{COOK20}], providing evidence for
dineutron correlations in these nuclei~[\cite{NAKA26}].
%We also showed the recent Coulomb dissociation measurement of $^8$He, where the four-neutron coincidence was also made in addition to the two-neutron coincidence. $\eone$ response up to the excitation energy of 12-14 MeV is 
%now possible, and the result shows the soft $\eone$ excitation arising from the $^6$He+$2n$ halo structure and the broader structure.

Coulomb dissociation has also been extensively applied to 
radiative-capture reactions.
Coulomb dissociation is the inverse process of 
radiative neutron or proton capture and provides a 
powerful indirect method for determining stellar reaction rates 
when direct radiative-capture measurements suffer 
from extremely small cross sections.
%kokomade 14:32 2026.8.8
An important advantage of Coulomb dissociation over direct radiative
capture is its much larger cross section, owing to the kinematic factor and the large number of virtual photons. In addition, performing the reaction at higher energies in inverse kinematics allows one to take advantage of kinematic focusing and the use of much thicker targets. 
Coulomb-dissociation studies of $^8$B and $^{15}$C are 
successful examples of this approach.

In the near future, we expect major progress in Coulomb-dissociation
studies both experimentally and theoretically.
Experimentally, as demonstrated for Coulomb dissociation of $^8$He, we are now able to measure up to four neutrons in coincidence. In the near future, we expect this to be extended to the coincident detection of up to six neutrons using new-generation neutron detectors such as NEOLITH at RIBF. 
For two-neutron halo nuclei, Dalitz-plot analyses can provide detailed information on three-body correlations. 
With greater acceptance, full $\eone$
strength spectra will be obtained up to the GDR region at higher-energy facilities. This will enable us to cover a much wider range of $\ex$
in a single experiment,  
from soft $\eone$ excitation through the PDR to GDR, 
as illustrated in Fig.~\ref{eoneres}.

A wider range of astrophysical reactions can be studied using Coulomb dissociation.
More precise measurements of angular distributions with much higher statistics, combined
with advanced reaction theories, will enable more reliable
determinations of stellar reaction rates.

We expect Coulomb dissociation to be applied more extensively 
to moderately neutron-rich nuclei
to investigate their pygmy dipole modes, 
which may provide important constraints
on the equation of state (EoS) of neutron-rich nuclear matter.
In medium-heavy nuclei, such as the Ca, Ni, and Sn isotopic chains, systematic studies of PDR are highly desirable, but experiments have so far
been limited to only a few cases~[\cite{WIEL09,WIEL18,ROSS13,ADRI05}].
This is attributed to the fact that the final states following Coulomb dissociation become more complex. The excited states can decay not only into a simple fragment plus a few neutrons, but also into multiple charged fragments, a few neutrons, and multiple $\gamma$ rays. 
With new-generation $\gamma$-ray detectors with nearly $4\pi$ acceptance being developed, we expect more systematic PDR studies to become possible.

Although Coulomb dissociation has become a mature and quantitatively reliable indirect method for investigating nuclear structure and radiative capture reactions, several important theoretical challenges remain. A major goal is the development of a unified microscopic reaction theory that consistently incorporates higher-order electromagnetic processes, nuclear interactions, continuum-continuum couplings, recoil, final-state interactions, and relativistic dynamics within a single framework. Progress is also needed toward fully relativistic formulations of the CDCC method and the dynamical eikonal approximation, particularly for experiments performed at several hundred MeV per nucleon and above. On the nuclear structure side, coupling reaction theories with modern \textit{ab initio} methods, chiral effective field theory, and energy-density-functional approaches would greatly reduce the dependence on phenomenological structure models and optical potentials. Additional challenges include the treatment of dynamic core excitation, three- and four-body breakup channels, nuclear-Coulomb interference, and proton-rich systems where Coulomb distortions play a dominant role. Finally, Bayesian uncertainty quantification, machine-learning emulators for computationally intensive reaction models, and time-dependent density-functional approaches offer promising avenues for improving the predictive power and reliability of Coulomb dissociation calculations. Addressing these challenges will further enhance the precision of Coulomb dissociation as a probe of exotic nuclei and as an indispensable indirect method to extract astrophysically important radiative capture cross sections.

Note that, as experimental data become more complex, increasingly sophisticated theoretical methods, such as those mentioned above, will be required for their interpretation. Further developments in dynamical and time-dependent calculations are also expected. Such approaches will improve our understanding of the electromagnetic response of exotic nuclei, including reaction dynamics, detailed three-body and many-body structures, and correlations such as $2n$, $4n$, and core-$n$ correlations, together with the relevant interactions in neutron-halo and neutron-skin nuclei. We also expect more reliable determinations of radiative-capture reaction rates.

In such Coulomb-dissociation studies, new-generation RI-beam facilities, such as RIBF at RIKEN, FRIB, FAIR, and HIAF, will play central roles.
Developments in experimental instrumentation, 
such as large-acceptance neutron-detector arrays and 
$\gamma$-ray detector arrays, are critical.
NEBULA~[\cite{NAKA16,KOND20}], NEBULA-plus at RIBF-RIKEN, MoNA~[\cite{BAUM05}] at MSU-FRIB and NEULAND~[\cite{BORE21}] at GSI-FAIR will play important roles.
We are currently developing a new-generation 
neutron-detector array, NEOLITH, which is capable of tracking recoil protons produced in plastic scintillators.
This will extend the capability for coincident neutron detection to up to six neutrons, with improved position and energy resolutions. 
Gamma-ray detector arrays such as CATANA~[\cite{TOGA20}] at SAMURAI/RIBF and CALIFA~[\cite{ALVA14}] at R3B/FAIR will be crucial for 
Coulomb-dissociation studies of heavier, moderately neutron-rich
nuclei and their PDRs.
In parallel to such experimental developments, 
developments in many-body and ab initio theories, as well as in dynamical reaction 
calculations, will also play crucial roles.
We also note that 
advances in high-performance computing are likely to improve our understanding of Coulomb dissociation processes 
and exotic nuclear states in the near future.

\section*{Acknowledgements}
C.A.B. acknowledges support from the U.S. Department of Energy under Grant No. DE-SC0026074.  
T.N. acknowledges support from JSPS KAKENHI Grant Nos. JP16H02179, JP18H05404, JP21H04465, and JP24H00006. C.A.B. and T.N. acknowledge support from the ExtreMe Matter Institute EMMI at the GSI Helmholtzzentrum f\"ur Schwerionenforschung.

%\begin{thebibliography}{99} 
%\bibliography{mainNotes.bib}% common bib file

\bibliographystyle{Harvard}
\bibliography{reference}

%\end{thebibliography}

\end{document}